\documentclass[letterpaper,11pt]{article}

\usepackage{natbib}
\usepackage{booktabs}
\usepackage{multirow}
\usepackage{tabularx} 
\usepackage{amsmath}  
\usepackage{amsfonts}
\usepackage{graphicx} 
\usepackage[margin=1in,letterpaper]{geometry} %
\usepackage{algorithm}
\usepackage{algorithmic}
\usepackage[final]{hyperref} 
\hypersetup{
	colorlinks=true,       
	linkcolor=blue,        
	citecolor=blue,        
	filecolor=magenta,     
	urlcolor=blue         
}

\newcommand*{\bv}[1]{\ensuremath \boldsymbol{#1}}

\newcommand*{\Xb}{\bv{X}}

\newcommand*{\xb}{\bv{x}}
\newcommand*{\yb}{\bv{y}}

\newcommand*{\xib}{\bv{\xi}}

\newcommand*{\thb}{\bv{\theta}}%
\newcommand{\Zb}{\bv{Z}}
\newcommand{\hW}{\hat{W}}
\newcommand{\hX}{\hat{x}}
\newcommand{\yt}{\bv{y}_{t_{1:m}}}
\newcommand{\It}{\bv{I}_{t_{1:m}}}

\providecommand{\myst}[1]{${#1}^{\textrm{st}}$}
\providecommand{\myrd}[1]{${#1}^{\textrm{rd}}$}
\providecommand{\nni}{\Delta^{\textrm{infec}}}
\providecommand{\ksens}{k_{\textrm{sens}}}
\providecommand{\kspec}{k_{\textrm{spec}}}
\renewcommand{\vec}[1]{\boldsymbol{{#1}}}
\providecommand{\tlock}{t_{\textrm{lock}}}

\begin{document}

\title{An approximate diffusion process for environmental stochasticity in infectious disease transmission modelling}
\author{Sanmitra Ghosh $^{1}$, Paul J. Birrell$^{2,1}$, Daniela De Angelis$^{1,2}$\\
\textit{\normalsize{$^{1}$MRC Biostatistics Unit, University of Cambridge, Cambridge, UK}}\\
\textit{\normalsize{$^{2}$UK Health Security Agency, London, UK}}
}
\date{}
\maketitle
\section*{Abstract}
Modelling the transmission dynamics of an infectious disease is a complex task. Not only it is difficult to accurately model the inherent non-stationarity and heterogeneity of transmission, but it is nearly impossible to describe, mechanistically, changes in extrinsic environmental factors including public behaviour and seasonal fluctuations. An elegant approach to capturing environmental stochasticity is to model the force of infection as a stochastic process. However, inference in this context requires solving a computationally expensive ``missing data" problem, using data-augmentation techniques. 
We propose to model the time-varying transmission-potential as an approximate diffusion process using a path-wise series expansion of Brownian motion. This approximation replaces the ``missing data" imputation step with the inference of the expansion coefficients: a simpler and computationally cheaper task. We illustrate the merit of this approach through two examples: modelling influenza using a canonical SIR model, and the modelling of COVID-19 pandemic using a multi-type SEIR model.

\section{Introduction}
\label{sec1}

Mathematical modelling of the complex dynamics of infectious diseases remains an essential tool to inform public health policies during epidemic outbreaks.
The major focus of such modelling work is describing the intrinsic transmission dynamics and the flow of individuals between compartments that segregate the population as per their disease state. However, an epidemic is also driven by a number of extrinsic factors,  including population mobility, social cycles (e.g. holidays), non-pharmaceutical interventions, and climatic variations  \citep{breto2009time}. In a compartmental model, such factors are often introduced explicitly through the description of the hazard (force) of infection when information about these external drivers is available \citep{knock2021key,keeling2020fitting,davies2020effects}.  However, while it is impossible to fully account for all extrinsic factors influencing transmission, yet ignoring this epistemic uncertainty often known as ``environmental stochasticity" leads to a structural miss-specification of the model, a ``model discrepancy". Model discrepancy can lead to miss-calibrated models that underestimate uncertainty and produce biased predictions \citep{Brynjarsd_ttir_2014}. An elegant approach to account for the un-modelled model discrepancy is to represent the force of infection as a stochastic process. For example, \cite{dureau2013capturing,cazelles2018accounting} use a diffusion process for this purpose, while \cite{birrell2021real} use a discrete time stochastic process. 
Parameter estimation for such stochastic models is, however, challenging. Inference, particularly in a Bayesian context, requires estimation of the joint posterior distribution of both the latent path of the stochastic process and the model parameters. Estimation using a Markov chain Monte Carlo (MCMC) algorithm, involves sampling the realisation of the stochastic process, a high dimensional object,  often through \textit{data-augmentation} techniques, which incur a hefty computational cost \citep{de2015four}. As a result efficient calibration of a compartmental model, which embeds a stochastic process, has received significant attention in the literature \citep[e.g.][]{fuchs2013inference,sottinen2008application} with the goal of alleviating the computational bottleneck associated with the inference of the stochastic process. 

In this paper we propose a new approach to the calibration problem through the use of a path-wise approximation of a diffusion process.
Specifically, we apply a truncated Fourier expansion of a Brownian motion to obtain the approximation. Application of this series expansion turns the task of inferring a high dimensional latent diffusion sample path into the task of inferring a smaller dimensional object, the expansion coefficients, which can be carried out without data-augmentation. This method is also applicable in the context of discrete time processes that converge to a diffusion in the continuous time limit. Such processes can be approximated by first carrying out the series expansion of the limiting diffusion and then applying a suitable time discretisation. We validate the proposed method against a data augmentation technique carried out using a particle MCMC sampler proposed in \cite{dureau2013capturing}, using a dataset from an influenza outbreak in a boarding school. We then apply this method to fit a model of COVID-19 spread in England during the first wave.

\section{Background: Epidemic models with a time-varying transmission-potential}\label{sec:Epidemic models with time-varying coefficients} 

We consider the canonical SIR (Susceptible-Infected-Removed) model \citep{anderson1992infectious} to introduce the stochastic modelling framework, although the methodology can be applied to other more complex compartmental models. 
In the SIR model the compartments denote the number of susceptible ($S$), infected ($I$), and recovered ($R$) people in a population subjected to an epidemic at time $t$. For a population of size $N$, the SIR model is defined by the following ODE system:
\begin{equation}\label{eq: basic SIR}
\frac{d S_t}{d t} = -\beta S_t \frac{I_t}{N}, \quad
\frac{d I_t}{d t}= \beta S_t \frac{I_t}{N} - \gamma I_t, \quad
\frac{d R_t}{d t} = \gamma I_t,
\end{equation}
where $\lambda=\beta \frac{I_t}{N}$ is the force of infection, describing the generation of infections with a transmission-potential $\beta$, between susceptible individuals and the fraction, $I_t /N$, of infectious individuals. The expected period spent in the compartment is given by $\gamma^{-1}$. The individual compartment sizes sum to $N = S_t + I_t + R_t$.

To include environmental stochasticity we introduce a time-varying $\beta_t$ \citep[e.g.][]{ellner1998noise,martinez2015unraveling,cauchemez2008likelihood,cauchemez2008estimating,cazelles1997using} 
  to mitigate model discrepancy, leading to a reformulation of the model in Eq~(\ref{eq: basic SIR}):
\begin{equation}\label{eq: basic SIR sde}
\begin{aligned}
dx_t &= a(x_t,\xib)dt + b(x_t,\xib) dW_t\\
\beta_t &= g(x_t)\\
\frac{d S_t}{d t} &= - \beta_t S_t \frac{I_t}{N}, \quad
\frac{d I_t}{d t}= \beta_t S_t \frac{I_t}{N} - \gamma I_t, \quad
\frac{d R_t}{d t} = \gamma I_t,
\end{aligned}
\end{equation}
where $x_t$ follows a diffusion process described by an It\^o stochastic differential equation (SDE) \citep{oksendal2013stochastic} with drift $a(\cdot)$, and diffusion $b(\cdot)$ functions parameterised by the vector $\xib$; $W_t$ is a standard Brownian motion; and $g(\cdot)$ is a nonlinear transformation that enforces $\beta_t>0$, such as exponential or inverse-logit transformation. Here we make some mild assumptions about $a(\cdot)$ and $b(\cdot)$ such as, for example, being locally Lipschitz with a linear growth bound \citep{oksendal2013stochastic} to ensure a non-explosive solution. 

Inference for the stochastic model in Eq~(\ref{eq: basic SIR sde}) within a Bayesian framework, requires inference of the latent sample path $\bv{x}$ of the diffusion $x_t$, which is indirectly observed through the time evolution of the disease states: $S_t,I_t,R_t$. This is a missing data problem that can be addressed through data-augmentation based MCMC methods \citep[e.g.][]{fuchs2013inference, dureau2013capturing} in which a high resolution (in time) Euler-Maruyama discretisation of $x_t$ is sampled along with the model parameters. Such MCMC methods incur high computational costs and have reduced efficiency in terms of mixing and speed of convergence. In what follows we will investigate a scalable approximation of $x_t$ that is faster to sample. 

\section{Methods}\label{sec:Series approximation of a diffusion process} 

Following \cite{LyonsSS12,luo2006wiener,pmlr-v151-ghosh22a}, we carry out a Fourier expansion of a Brownian motion $W_t$ and obtain a smooth path-wise series approximation. Using this approximation of a Brownian motion, we can in turn approximate the SDE for $x_t$ with a random ODE. Inference of $x_t$ can then be carried out by inferring coefficients of this ODE, without requiring data-augmentation.

\subsection{Fourier expansion of Brownian motion}\label{sec:Fourier expansion of Brownian motion} 

Within a time interval $[0,T]$, where $T$ is the length of the time horizon within which an epidemic is analysed, the Fourier expansion of a Brownian motion $W_t$ is given by \citep{luo2006wiener}:
\begin{equation}\label{eq: Brownian series}
    W_t = \sum_{i=1}^\infty\Big(\int_{0}^T\phi_i(s)dW_s\Big)\int_{0}^t\phi_i(u)du.
\end{equation}
where $\{\phi_i\}_{i=1}^{\infty}$ is a complete orthonormal basis of $L^2[0,T]$ (see Appendix A for derivation). For example this can be the generalised Fourier cosine basis \citep{lyons2014series} given by 
\begin{equation}\label{eq: KL}
    \phi_i(t)=( 2/T)^{1/2}\cos\{(2i-1)\pi t/2T\}.
\end{equation}
We will use the shorthand $Z_i=\int_{0}^T\phi_i(s)dW_s$. Since the basis functions $\{\phi_i\}$ are deterministic and orthonormal, it follows from standard results of It\^o calculus that $Z_i\sim \mathcal{N}(0,1)$ \citep{luo2006wiener}. By truncating the infinite series in Eq~(\ref{eq: Brownian series}) to $n$-terms we obtain a path-wise approximation of the Brownian motion $W_t$ given by 
\begin{equation}\label{eq:approximate Brown}
    \hat{W}_t = \sum_{i=1}^n Z_i\int_{0}^t\phi_i(u)du.
\end{equation}

\subsection{Approximating a SDE with a random ODE}\label{sec:Approximating a SDE with a random ODE} 

Taking derivative of $\hat{W}_t$ with respect to time we obtain the following approximation to white noise, the derivative of Brownian motion, given by
\begin{equation}\label{Wong-Zakai}
    \frac{d\hW_t}{dt}=\sum_{i=1}^n Z_i \phi_i(t).
\end{equation}

Now, let us replace the It\^o SDE in Eq~(\ref{eq: basic SIR sde}) with the following Stratonovich SDE \citep{oksendal2013stochastic}
\begin{equation}\label{eq: strat}
    dx_t = a'(x_t,\xib)dt + b(x_t,\xib) \circ dW_t,
\end{equation}
where $(\circ)$ denotes a Stratonovich integral \citep{oksendal2013stochastic} with respect to $W_t$. The It\^o SDE in Eq~(\ref{eq: basic SIR sde}) and the Stratonovich SDE given above are equivalent \citep{oksendal2013stochastic} if
\begin{equation}\label{eq: ito-strat}
    a'(x_t,\xib) = a(x_t,\xib) - \frac{b(x_t,\xib)}{2}\frac{\partial b(x_t,\xib)}{\partial x_t}b(x_t,\xib).
\end{equation}
By substituting the term $dW_t$ in Eq~(\ref{eq: strat}) with the approximation $d\hW_t$ in Eq~(\ref{Wong-Zakai}), we obtain the following (random) ODE: 
\begin{equation}\label{eq: strat ode}
    \frac{d\hX_t}{dt}=a'(\hX_t,\xib) + b(\hX_t,\xib)\sum_{i=1}^n Z_i \phi_i(t).
\end{equation}
The work of \cite{wongzakai} shows that as $n\rightarrow\infty$ the solution  $\hX_t$ of the above ODE will converge to the solution $x_t$ of the Stratonovich SDE Eq~(\ref{eq: strat}) which, given the choice of $a'(\cdot)$ in Eq~(\ref{eq: ito-strat}), is an equivalent representation of the It\^o SDE in  Eq~(\ref{eq: basic SIR sde}). Thus, the series approximation $\hX_t$ of the solution $x_t$ of an It\^o SDE converges to the solution of an equivalent Stratonovich SDE.

Next, we discuss the implications of the above approximation with regards to inference.

\subsection{Inference using the series approximation}\label{sec:Inference using the series approximation}

Using the path-wise series approximation of a diffusion process $x_t$, presented in the previous sections, we can re-write the canonical SIR model in Eq~(\ref{eq: basic SIR sde}) as a system of coupled ODEs given by
\begin{equation}\label{eq: SIR fourier}
\begin{aligned}
\frac{d\hX_t}{dt} &=a'(\hX_t,\xib) + b(\hX_t,\xib)\sum_{i=1}^n Z_i \phi_i(t)\\
\beta_t &= g(\hX)\\
\frac{d S_t}{d t} &= - \beta_t S_t \frac{I_t}{N}, \quad
\frac{d I_t}{d t}= \beta_t S_t \frac{I_t}{N} - \gamma I_t, \quad
\frac{d R_t}{d t} = \gamma I_t,
\end{aligned}
\end{equation}
where $a'(\cdot)$ is given by Eq~(\ref{eq: ito-strat}).
Note that the randomness in the above model is now encapsulated in the expansion coefficients $\Zb =(Z_1,\ldots,Z_n)$. Inference in this model is then relegated to the inference of all the parameters: $\Zb, \xib, \gamma$, and the initial values: $x_0,S_0,I_0,R_0$. We denote the vector of the parameters governing the dynamics as $\thb=(\xib, \gamma)$. We denote the state vector evolving in continuous time by $\Xb_t = (x_t,S_t,I_t,R_t)$, and by $\Xb_0 =(x_0,S_0,I_0,R_0)$ the vector of initial values.

In order to explain the inferential framework based on the series approximation, in Eq~(\ref{eq: SIR fourier}), we assume that the available data $\yt=(y_{t_{1}},\ldots, y_{t_{m}})$ are the noisy observations of the state $I_t$ at $m$ time-points. Here we are simply considering prevalence data for the ease of exposition, however the same idea can be extended to more complex observational models where the observed data only provide partial (and often indirect) information of the states $\Xb_t$ \citep{birrell2021real}. 

The inferential goal is to learn the posterior distribution of all the unknown quantities, given the data $\yt$. We place priors $p(\thb)$, $p(\Zb)$, $p(\Xb_0)$ on the parameters, expansion coefficients and the initial values. Note that, by construction, the $\Zb = (Z_1, \ldots, Z_n)$ have an independent standard Normal prior, see Section \ref{sec:Fourier expansion of Brownian motion}. We then numerically solve Eq~(\ref{eq: SIR fourier}) to obtain a likelihood $p(\yt|\It, \bv{\epsilon})$, based on the noise assumption, where $\It$ is the numerical solution of the state $I_t$ evaluated at the $m$ time-points, and $\bv{\epsilon}$ are the parameters of the chosen data distribution. The posterior distribution, up to a normalisation constant, follows from the Bayes rule:
\begin{equation}\label{eq: pos ode approx diff}
   p(\thb,\Zb,\Xb_0|\yt) \propto p(\yt|\It, \bv{\epsilon})p(\thb)p(\Zb)p(\Xb_0)p(\bv{\epsilon}).
\end{equation}
Samples from the posterior distribution can be obtained using MCMC. The samples of the latent approximate diffusion path $\hat{\xb}$ are simply the numerical solution of the ODE for $\hX_t$ evaluated using samples of $\thb,\Zb,\Xb_0$ from the posterior distribution.

Note that if we had described $\beta_t$ using a SDE, then to sample the latent diffusion $\xb$ we would have had to use data-augmentation. This involves imputing the sample path of the latent diffusion at the time-points of observations $t_{1:m}$ as well as at time-points in-between the observations using, say, the Euler-Maruyama scheme \citep{kloedenplatenbook}. If one chooses $l$ time-points between $t_m$ and $t_{m-1}$ then the MCMC sampler would target $m(l+1) - l$ random variables (including $x_0$) related to the diffusion. Using the proposed approximation we have replaced the inference of $m(l+1) - l$ variables with $n$, which is a simpler inference problem if $n<m(l+1) - l$. Below we show that choosing a value of $n$ substantially smaller than $m(l+1) - l$ still renders an estimate of the posterior distribution that is a reliable approximation to the true posterior. 

\section{Evaluation}\label{sec:Evaluation}

To evaluate the proposed approximation method we fit the model in Eq~(\ref{eq: SIR fourier}) to the data of an outbreak of influenza at a  boarding school \citep{jackson2013school} (see Fig~\ref{ouppc compare real} (a)), on the number of infections for a period of $T=14$ days among a population of size $N = 763$. This dataset is publicly available in the R package \texttt{outbreaks} \citep{ob}. This dataset was previously used in \cite{del2015sequential,ryder18a} to fit a SIR model under assumption that the time varying transmission-potential can be modelled using an Ornstein–Uhlenbeck (OU) process. The model in \cite{del2015sequential} is similar to the stochastic model introduced in Eq~(\ref{eq: basic SIR sde}). Using the OU SDE for $x_t$ we can write the model in Eq~(\ref{eq: basic SIR sde}) as:
\begin{equation}\label{eq: SIR OU}
\begin{aligned}
dx_t &= (\xi_1 - \xi_2 x_t)dt +  \xi_3 dW_t \\
\beta_t &= \exp(x_t)\\
\frac{d S_t}{d t} &= - \beta_t S_t \frac{I_t}{N}, \quad
\frac{d I_t}{d t}= \beta_t S_t \frac{I_t}{N} - \gamma I_t, \quad
\frac{d R_t}{d t} = \gamma I_t,
\end{aligned}
\end{equation}
where $\xib=(\xi_1,\ldots, \xi_3)$ denotes the parameter vector of the OU SDE. 

Here we specifically want to compare the outcome of inference using the true OU diffusion used above (\textbf{SDE}) with its series approximation (\textbf{SA}), leading to a model such as in Eq~(\ref{eq: SIR fourier}), given by
\begin{equation}\label{eq:OU series}
    \frac{d\hX_t}{dt} =(\xi_1 - \xi_2 x_t) + \xi_3\sum_{i=1}^n Z_i\phi_i(t),
\end{equation}
where we have chosen the generalised Fourier basis Eq~(\ref{eq: KL}) as the function $\phi_i(t)$. 

For the \textbf{SDE} model the latent sample path $\bv{x}$, the diffusion parameters $\xib$, initial value $x_0$ and the parameter $\gamma$ were also estimated together with the initial susceptibility, $s_0 = S(t=0)/N$, assuming the initial recovered fraction $r_0=0$ and thus $i_0 = 1 - s_0$. As this is count data we have specified a Poisson likelihood:
\begin{equation}
   y_{t_i}|\thb, \bv{x}, \Xb_0 \sim \operatorname{Poisson}(I_{t_i}), \quad i=1, \ldots, m,
\end{equation} 
where in this case $\Xb_0=(x_0,s_0)$. For the \textbf{SA} model we used the inferential framework introduced in the previous section and used the Poisson likelihood as above:
\begin{equation}
     y_{t_i}|\thb, \Zb, \Xb_0 \sim \operatorname{Poisson}(I_{t_i}).
\end{equation}

We chose a weakly-informative prior for the parameters governing the dynamics $\xi_1,\ldots, \xi_3, \gamma \sim \Gamma(2,2)$. For $s_0$ a $\operatorname{Beta}(2,1)$, since we expect the true value to be near or greater than $2/3$, and for the initial value of the diffusion we used a prior $x_0 \sim \mathcal{N}\Big(\xi_1/\xi_2,\big(\frac{\xi^2_3}{2 \xi_2}\big)^2\Big)$, which is the stationary distribution of the OU diffusion.

For the \textbf{SDE} model, data-augmentation using a particle filter was employed to sample the `true' diffusion's path, following \cite{dureau2013capturing}, and produce an unbiased estimate of the likelihood. Parameters $\gamma,\xib,\Xb_0$ were estimated using the Metropolis-Hastings (MH) algorithm, with an adaptive random-walk proposal based on algorithm 4 of \cite{andrieu2008tutorial}. See  B in S1 text for further details on this proposal mechanism. The likelihood estimate produced by the particle filter was used in the acceptance step of the MH algorithm. This particle-marginal Metropolis-Hastings (PMMH) MCMC scheme for jointly updating the latent diffusion path along with the parameters has been shown to have superior performance \citep{dureau2013capturing} when compared to other data-augmentation approaches. For the PMMH, we used a Bootstrap particle filter \citep{gordon1995bayesian}, where the particles are propagated using Euler-Maruyama discretisation, and set the number of particles to $1000$. Following \cite{del2015sequential}, we carried out the Euler-Maruyama iterations with a stepsize $\delta t = 0.1$, leading to $l=9$ time-points between two observations.

For the \textbf{SA} model we used the Metropolis-Hastings algorithm with the same adaptive random-walk proposal (RWMH) used with the PMMH scheme and the Euler method to numerically solve the ODE adopting the same step-size that is used with the Euler-Maruyama scheme for the SDE.

Note that inference for the \textbf{SDE} model using PMMH will be substantially more computationally heavy compared to the inference for the ODE based \textbf{SA} model,  irrespective of the value of $n$. This is due to  the particle filter requiring multiple evaluation of the Euler-Maruyama scheme at each MCMC iteration. Even when parallelised, the particle filter will be bottlenecked by a weight-updating step (see \cite{gordon1995bayesian} for details) requiring message-passing across processes. The Euler scheme for solving the ODE in Eq~(\ref{eq:OU series}), in comparison, is evaluated once every iteration of a Metropolis-Hastings algorithm targeting the posterior distribution in Eq~(\ref{eq: pos ode approx diff}). 

\begin{figure*}[!ht]
  \centering
\includegraphics[width=0.7\textwidth,keepaspectratio=true]{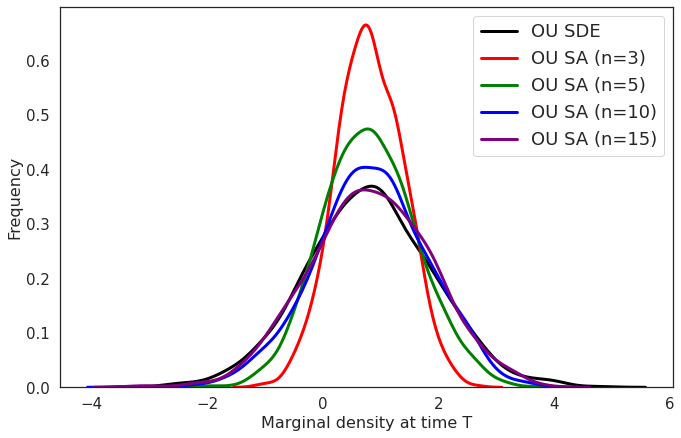}
\caption{Comparison between the marginal density of the OU SDE at time $T=14$, with that obtained through the series approximation upon varying the number of basis $n=3,5,10,15$.}
\label{Figure:wv compare}
\end{figure*}

A crucial parameter for the proposed method is the number of basis functions $n$. If a value of $n$ produces a close match between the marginal densities of the true and approximate diffusion at the end of the analysis period $T$ then the approximation will be valid throughout the course of the epidemic. In this case $T=14$. In Fig~\ref{Figure:wv compare} we compare the time $T$ marginal densities $p(\hat{x}_t)|_{t=T}$ obtained by solving the ODE in Eq~(\ref{eq:OU series}) associated with the \textbf{SA}, and $p(x_t)|_{t=T}$ obtained from the original OU diffusion, both based on some trial parameters sampled from the prior. The value $n=15$ produces a close match  between the marginal densities. We defer further discussion of the effect of $n$ on estimation to section \ref{sec:Experiment 2}.

\subsection{Results: comparison between true and approximate diffusion}\label{sec:Experiments}

We fitted the two models, \textbf{SDE} and \textbf{SA} respectively, using the associated algorithms as described above to the influenza dataset. We ran two chains of both PMMH, for the \textbf{SDE} model, and RWMH, for the \textbf{SA} one, for $10^6$ iterations where the first $5 \times 10^5$ iterations were discarded as burnin and the remaining samples thinned to obtain $1000$ samples from the posterior distribution. The running times were $15907$ and $2397$ seconds for the PMMH and RWMH with $n=15$, respectively. We implemented a vectorised particle filter and the Euler solver for the ODE using \texttt{Jax} \citep{jax2018github}. The adaptive MCMC algorithm was implemented using \texttt{Python}.

\begin{figure}[!ht]
\includegraphics[width=\textwidth,keepaspectratio=true]{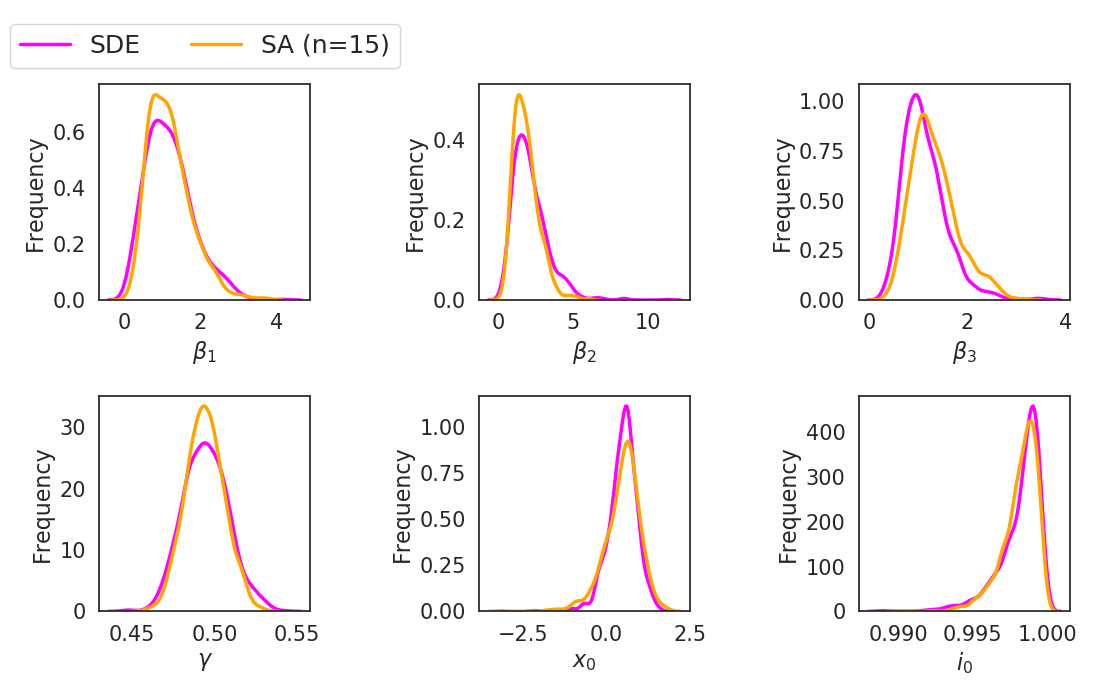}
\caption{Comparison of the posterior marginal densities of the parameters obtained using the \textbf{SDE} and the \textbf{SA} (with $n=15$ basis function). These densities are summarised using a kernel density estimate.}
\label{oumarg compare real}
\end{figure}

\begin{figure}[!ht]
 \includegraphics[width=\textwidth,keepaspectratio=true]{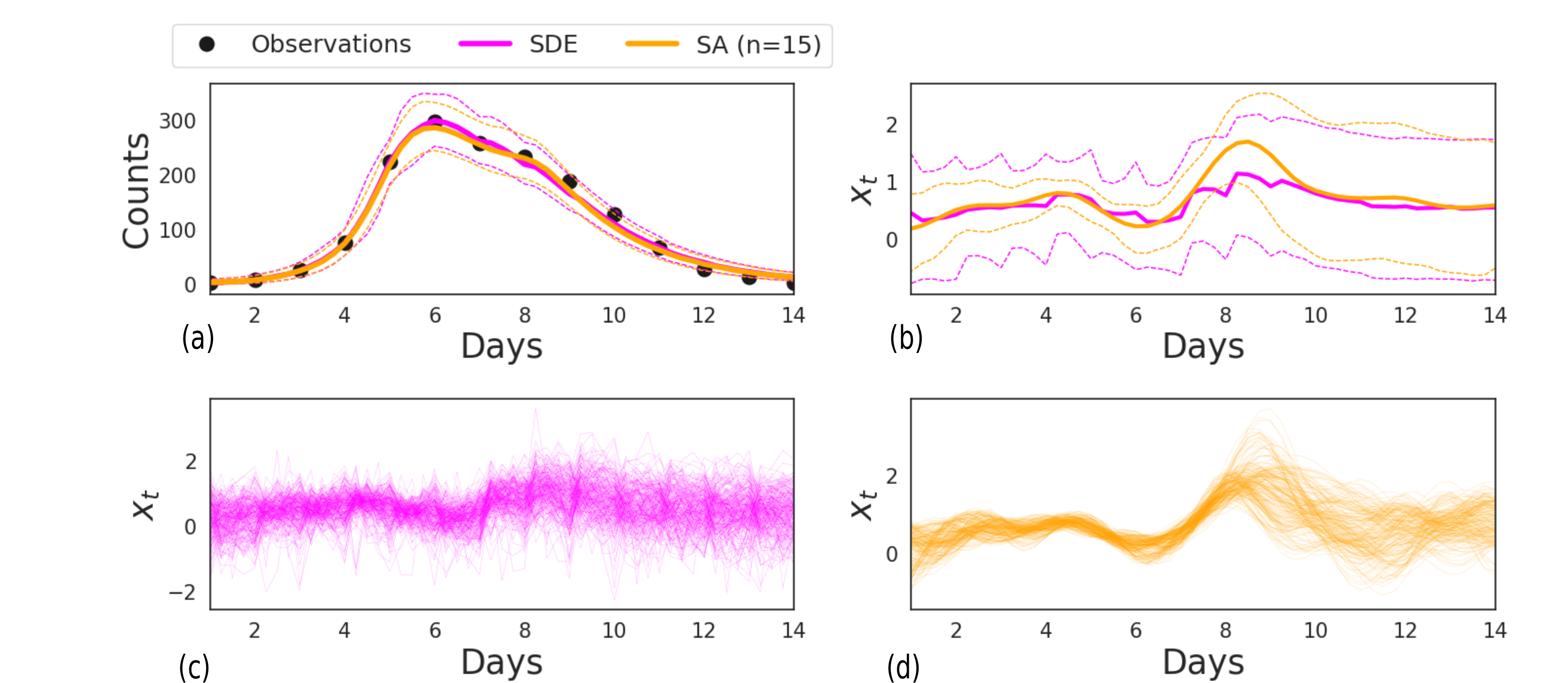}
\caption{\textbf{Influenza dataset}: Goodness-of-fit (a); posterior distribution of the latent diffusion paths corresponding to the \textbf{SDE} and \textbf{SA} counterparts (b), with densities summarised by the mean (solid lines) and $95\%$ credible intervals (broken lines); and samples from the posterior distribution of the latent diffusion paths, \textbf{SDE} (c) and \textbf{SA} (d)}
\label{ouppc compare real}
\end{figure}

We notice a good agreement between the parameter estimates obtained using the \textbf{SDE} and \textbf{SA} counterparts (see Fig~\ref{oumarg compare real}). Furthermore, in Fig~\ref{ouppc compare real} we compare the goodness-of-fit and display the posterior distribution of the latent diffusion paths $p(\bv{x}|\yt)$ and $p(\hat{\bv{x}}|\yt)$, corresponding to the \textbf{SDE} and \textbf{SA}. Additionally, for aid of visualisation, we have also plotted draws from the (posterior) sample paths for both models in Fig~\ref{ouppc compare real} (c) and (d).

We observe a good agreement between the epidemic curves obtained using the \textbf{SDE} and the \textbf{SA}, but for the posterior distribution of the latent diffusion paths the credible intervals are narrower for the \textbf{SA}. The \textbf{SA}, due to the truncation of the infinite series expansion, produces smoother paths, slightly underestimating the volatility of the latent diffusion path. On a closer introspection of the posterior means (Fig~\ref{ouppc compare real} (b)), it is noticeable that the latent diffusion paths drop and increase again in the period between the $4$-th and $9$-th day, around the peak, indicating sudden changes in the transmission-potential. These changes are reflected in the estimates of both \textbf{SDE} and \textbf{SA}. After the $9$-th day, the variability in the latent paths increase for both \textbf{SDE} and \textbf{SA} and the posterior means match closely. This is expected since after the peak, when the epidemic is receding, a large change in $\beta_t$ will have negligible effect on the case counts. 

These results were confirmed in a simulation study where the simulated datasets mimicked this influenza dataset (see Appendix C). 

\subsection{Sensitivity to the choice of $n$}\label{sec:Experiment 2}

In Fig~\ref{Figure:wv compare} we noticed that the marginal distribution of the latent diffusion path and its series approximation starts agreeing beyond $n\geq10$ terms. It is worth investigating whether such a threshold exist for the posterior distributions obtained using the \textbf{SDE} and the \textbf{SA}. We did this by further comparing the joint posterior distribution $p(\thb,\Xb_0|\yb)$, from \textbf{SDE} and \textbf{SA} while varying $n$. Note that $\thb$ and $\Xb_0$ are quantities which were estimated using both the \textbf{SDE} and \textbf{SA} counterparts, and thus the joint posterior of these were chosen for comparison. For this comparison we estimated the posterior distribution by fitting the \textbf{SA} repeatedly with number of basis set to $n=3,5,10,15,20,25,30$. To compare the posterior distributions, we used the maximum mean discrepancy (MMD) metric \citep{gretton2012kernel}, a divergence metric that can be calculated using samples from the distributions. See Appendix H for further details on this metric.

In Fig~\ref{oummd compare} we plot the MMD between the posteriors from \textbf{SDE} and \textbf{SA} for increasing $n$. For $n \geq 10$ we found good agreement between the two posteriors, consistent with the results from comparing the marginal densities (Fig~\ref{Figure:wv compare}). This reinforces our approach of choosing the number of basis by comparing marginals of the latent process, while using the \textbf{SA}. We summarise the runtimes of MCMC with the RWMH proposal for each choice of $n$ in Table~\ref{runtimes}, noting that the increase in the runtimes as we varied $n$ was negligible, especially when compared to the PMMH with \textbf{SDE}. 

\begin{figure}[!ht]
\centering
\includegraphics[width=0.7\textwidth,keepaspectratio=true]{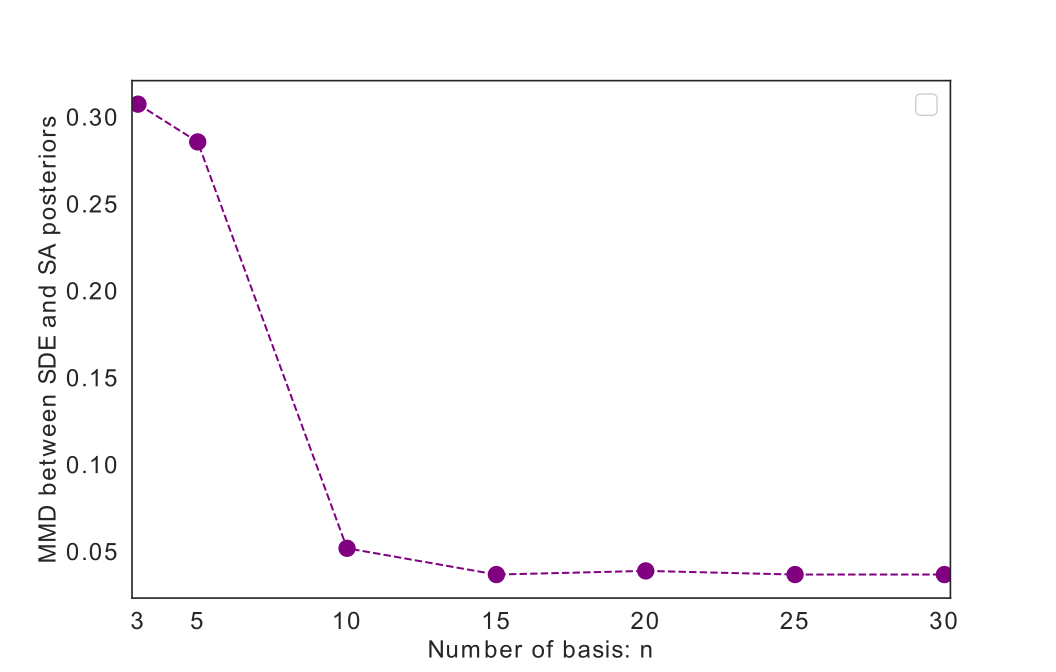}
\caption{MMD between the joint posterior distributions of the parameters $\thb$ and initial values $\Xb_0$ from \textbf{SDE} and \textbf{SA} (for different $n$).}
\label{oummd compare}
\end{figure}

\begin{table}[t]
\caption{\bf Runtimes (rounded to nearest integer), in seconds, of MCMC for \textbf{SA}, as a function of the number of basis $n$, in comparison with the runtime of \textbf{SDE}. These were run on a $3.6$ GHz machine with $16$ GB memory.}
\label{runtimes}
\begin{center}

\begin{tabular}{|ll|l|l|l|l|l|l|l|}
\hline
\multicolumn{9}{|c|}{\bf Runtimes in seconds} \\
\hline
\textbf{SA} with & $n=3$ & $n=5$ & $n=10$ & $n=15$ & $n=20$ & $n=25$ & $n=30$ & \textbf{SDE}\\
\hline
& $2280$  &$2282$ &  $2337$ & $2397$ & $2470$ & $2538$ & $2637$ & $15907$\\
\hline
\end{tabular}

\end{center}
\end{table}

\section{Application: modelling COVID-19 outbreak in England}\label{sec:Application: modelling COVID-19 outbreak in England}

Our proposed method of modelling the time-varying transmission-potential as an approximate diffusion can also be applied to a discrete time stochastic process that converges to a diffusion in the continuous time limit. For example, an AR$(1)$ process converges to a OU diffusion. Thus, if one is already using an AR$(1)$ process to model the transmission-potential, then a discretised version of the series approximation of OU diffusion, the ODE in Eq~(\ref{eq:OU series}), can be chosen as its replacement. 

To exemplify the application of the series expansion method in replacing a discrete time stochastic process, we have chosen to fit a compartmental model whose dynamics are described as a set of first order difference equations, to data from the first wave of the COVID-19 outbreak in England, between February and August 2020 \citep{birrell2021real}. This model captures the effect of unknown extrinsic factors on the force of infection through a time-varying transmission-potential modelled as a Gaussian random-walk. We introduce the model of \cite{birrell2021real} in what follows and introduce an alternative formulation using the series approximation of Brownian motion.

\subsection{Transmission model for COVID-19}\label{sec:The transmission model}

This is an age and spatially structured transmission model, stratifying the population into $n_A=7$ age groups and $n_r=7$ regions. Within each region, the transmission dynamics are governed by a system of first order difference equations:
\begin{equation}\label{eqn:determ.dynam}
\begin{aligned}
 S_{r, t_k, a} &= S_{r, t_{k - 1}, a}\left(1 - \lambda_{r, t_{k - 1}, a}\delta t\right)\\
 E^1_{r, t_k, a} &= E^1_{r, t_{k - 1}, a}\left(1 - \frac{2\delta t}{d_L}\right) + S_{r, t_{k - 1}, a}\lambda_{r, t_{k - 1}, a}\delta t\\
 E^2_{r, t_k, a} &= E^2_{r, t_{k - 1}, a}\left(1 - \frac{2\delta t}{d_L}\right) + E^1_{r, t_{k - 1}, a}\frac{2\delta t}{d_L}\\
 I^1_{r, t_k, a} &= I^1_{r, t_{k - 1}, a}\left(1 - \frac{2\delta t}{d_I}\right) + E^2_{r, t_{k - 1}, a}\frac{2\delta t}{d_L}\\
 I^2_{r, t_k, a} &= I^2_{r, t_{k - 1}, a}\left(1 - \frac{2\delta t}{d_I}\right) + I^1_{r, t_{k - 1}, a}\frac{2\delta t}{d_I}\\
\end{aligned}
\end{equation}
where: $S_{r, t_k, a}$, $E^d_{r, t_k, a}$, $I^d_{r, t_k, a}, d = 1, 2$ represent the time $t_k, k = 1, \ldots, K,$ partitioning of the population of individuals in a region $r, r = 1, \ldots, n_r$, in age-group $a, a = 1, \ldots, n_A$, into $S$ (susceptible), $E$ (exposed) and $I$ (infectious) disease states. The average period spent in the exposed and infectious states are given by the parameters $d_L$ and $d_I$ respectively; and $\lambda_{r,t_k,i}$ is the time- and age-varying rate with which susceptible individuals become infected, the force of infection. Time steps of $\delta t = 0.5$ days are chosen to be sufficiently small relative to the latent and infectious periods. Following \cite{birrell2011bayesian} the initial conditions of the system states $S,E^1,E^2,I^1,I^2$ at $t_0$ are given by region-specific parameters $\psi_r$ and $I_{0,r}$, describing the initial exponential growth and the initial number of infectious individuals, respectively. New infections are generated as
  \begin{equation}\label{eq:nni}
    \nni_{r,t_k,a} = S_{r,t_k,a}\lambda_{r,t_k,a} \delta t,
  \end{equation} 
where $\lambda_{r,t_k,a} \delta t$ is driven over time by a region-specific time-varying potential, $\beta_{t_k,r}$, which moderates the rate at which effective contact take place. This region-specific transmission-potential captures the discrepancy between how actual contact take place between the age groups, and that encoded by a set of time-varying contact matrices. We refer the reader to \cite{birrell2021real} for further details on the model dynamics and parameterisation. 

Over time $\beta_{t_k,r}$ is not allowed to vary unconstrained and a smoothing is imposed by assuming, {\it a priori} that its evolution follows a Gaussian random-walk process with volatility $\sigma_{\beta_{t}}$: 
\begin{equation}\label{eq: random ealk beta fast}
\begin{aligned}
      \log\left(\beta_{t_k,r}\right) &\sim  \mathcal{N}\left(\log\left(\beta_{t_{k-1},r}\right),\sigma_{\beta_{t}}^2\right), \quad \text{if $t_k > \tlock$},\\
      \log\left(\beta_{t_k,r}\right) &= 0, \quad \text{if $t_k \leq \tlock$},
\end{aligned}
\end{equation}
where $\tlock$ indicates the time-point corresponding to the lockdown introduced in England on \myrd{23} March $2020$. This random-walk formulation requires the inference of the high-dimensional (due to the choice of $\delta t$) sample path of this process, an extremely challenging task using MCMC. We will discuss this inferential difficulty later in Section \ref{sec: Challenge of sampling the latent contact-rate: an alternative formulation}. To restrict the dimensionality of the process, in \cite{birrell2021real} this transmission-potential is assumed to be piecewise constant with weekly changepoints, and its values at these changepoints modelled as a random-walk. Denote $w_k \equiv w(t_k)$ the week in which time $t_k$ falls. Then the time evolution of the transmission-potential is modelled at a slower weekly time-scale:
\begin{equation}\label{eq: random ealk beta}
\begin{aligned}
      \log\left(\beta_{w_k,r}\right) &\sim  \mathcal{N}\left(\log\left(\beta_{w_{k-1},r}\right),\sigma_{\beta_{w}}^2\right), \quad \text{if $t_k > \tlock + 7/\delta t$},\\
      \log\left(\beta_{w_k,r}\right) &= 0, \quad \text{if $t_k \leq \tlock+ 7/\delta t$},\\
       \beta_{t_k,r} &= \beta_{w_k,r},
\end{aligned}
\end{equation}
as a Gaussian random-walk, with volatility $\sigma_{\beta_{w}}$, following the first week of the lockdown. Realisation of the process, for each region, can then be obtained by sampling the vector $\Delta \bv{\beta}_r$ of all the weekly increments $\Delta\beta_{w_{k},r}=\log\left(\beta_{w_{k},r}\right)-\log\left(\beta_{w_{k-1},r}\right)$. 
It was assumed in \cite{birrell2021real} that the contact matrices sufficiently described how actual contacts took place between different age groups prior to the lockdown and thus $\beta_{w_k,r} = 1$ over that period.

\subsection{Inference}

To fit the model, using a Bayesian framework, surveillance data of age- and region-specific counts of deaths in people with a lab-confirmed COVID-19 diagnosis between \myst{17} February and \myst{1} August was used. Furthermore, serological data from NHS Blood and Transplant (NHSBT), informing the fraction of the population carrying COVID-19 antibodies, were also used.

Following \cite{birrell2021real} the number of observed deaths $y^d_{r,t_k,a}$ on day $t_k$, in age group $a$, and in region $r$ follows a negative binomial distribution:
\begin{equation}
  y^d_{r,t_k,a}|d_I,p_a,\psi_r,I_{0_{r}},\Delta \bv{\beta}_r,\eta \sim \textrm{NegBin}\left(\mu_{r,t_k,a}, \eta\right),
\end{equation}
where the mean $ \mu_{r,t_k,a} = p_a \sum_{l=0}^k f_{k - l} \nni_{r, t_l, a}$ is derived using Eq~(\ref{eq:nni}), an assumed-known distribution of the time from infection to death from COVID-19, $f$, and an age-specific infection-fatality ratio $p_a$. Here $\eta$ is a dispersion parameter such that $\mathbb{E}y^d_{r,t_k,a} = \mu_{r,t_k,a}$ and $\textrm{Var}\left(y^d_{r,t_k,a}\right) = \mu_{r,t_k,a}\left(1 + \eta\right)$.

If, on day $t_k$, $n_{r,t_k,a}$ blood samples are taken from individuals in region $r$ and age-group $a$, and the observed number of positive tests is $y^s_{r,t_k,a}$, then
\begin{equation}
  y^s_{r,t_k,a}|d_I,\psi_r,I_{0_{r}},\Delta \bv{\beta}_r,\ksens,\kspec   \sim \textrm{Bin}\left(n_{r,t_k,a}, \ksens \left(1 - \frac{S_{r,t_k,a}}{N_{r,a}}\right) + \left(1 - \kspec\right)\frac{S_{r,t_k,a}}{N_{r,a}}\right),
\end{equation}
where $\ksens$ and $\kspec$ parametrises the sensitivity and the specificity of the serological testing process, and $S_{r,t_k,a}$ is obtained by solving the difference equations in Eq~(\ref{eqn:determ.dynam}). $N_{r,a}$ is the total population in age-group $a$ and region $r$.

The unknown quantities that need to be inferred can be divided into two groups: (i) Global parameters $\thb_g = (\eta,d_I,p_1,\ldots,p_{n_{A}},\ksens,\kspec,\sigma_{\beta_{w}})$ shared between regions, and (ii) regional parameters specific to each region:  $\thb_r = (\psi_r,I_{0_{r}},\Delta \bv{\beta}_r)$. After placing the same priors as was used in \cite{birrell2021real} (and listed in Appendix E), the posterior distribution of the unknown quantities is as follows:
\begin{equation}\label{eq:posterior}
    p(\thb_g,\thb_1, \ldots, \thb_{n_{r}}|\bv{y^d},\bv{y^s})\propto p(\thb_g)\prod_{k=1}^K \prod_{a=1}^{n_{A}} \prod_{r=1}^{n_{r}} p(y^d_{r,t_k,a}|\thb_g,\thb_r)p(y^s_{r,t_k,a}|\thb_g,\thb_r)p(\thb_r),
\end{equation}
where we denote by $\bv{y^d},\bv{y^s}\in \mathbb{R}^{K \times {n_A} \times n_{r}}$ the data for all time-points, ages and regions corresponding to deaths and sero-positive tests, respectively. 

\subsubsection{Sampling from the posterior}

Sampling from the posterior distribution Eq~(\ref{eq:posterior}) is challenging due to the large number of random-walk increments corresponding to all regions and weeks since lockdown. MCMC with a vanilla RWMH proposal, as applied in \cite{birrell2021real}, due to the linear scaling of convergence time with increasing dimensions 
mixes poorly and requires a large number of iterations ($\approx 10^7$) of the Markov chain before convergence is reached. To improve convergence we instead used a random-scan Metropolis-within-Gibbs (MwG) algorithm that circumvent the updating of a large parameter vector at each iteration. This MwG algorithm exploits the independence between the regional parameters. Our proposed sampling strategy consists of sampling alternatively, at each MCMC iteration, from the posterior of the global parameters conditioned on all the regional ones: (i) $p(\thb_g|\thb_1,\ldots,\thb_{n_{r}},\bv{y^d},\bv{y^s})$, and regional parameters for one randomly chosen region conditioned on the global ones (since the regional parameters are conditionally independent of any other region's parameters): (ii) $p(\thb_{r^{*}}|\thb_g,\bv{y^d},\bv{y^s})$, where $r^{*}\sim \operatorname{Uniform}(1,n_r)$. Samples from each of these conditional distributions are obtained using an adaptive RWMH move with the same adaptation mechanism used in Section \ref{sec:Experiments}. The pseudocode for this MwG algorithm is furnished in Appendix F.

\subsubsection{An alternative formulation}\label{sec: Challenge of sampling the latent contact-rate: an alternative formulation}

The number of region-specific random-walk increments $\Delta\beta_{w_{k},r}$ that needs to be sampled increases with time. The performance of the MwG algorithm starts deteriorating and exhibiting poor mixing and slow convergence, as this number becomes large. This limits dramatically the usefulness of this model in the context of a real-time application. 

 For the model in Eq~(\ref{eq: random ealk beta}), this problem can be tackled by increasing the time between two successive changepoints thus reducinge the number of increments to be sampled for a period of analysis. This is however driven by computational convenience, and it would be more meaningful to learn these changes from data. We could model the time evolution of the transmission-potential at a faster time-scale, for example as in Eq~(\ref{eq: random ealk beta fast}). However, in this case the number of random-walk increments, to be sampled per region, equals the number of time-points between lockdown and the end of analysis date. Any MCMC sampler, that uses a RWMH proposal, would struggle severely to move efficiently in such a high-dimensional parameter space. 

To alleviate these problems we propose to model the transmission-potential as a Brownian motion $W_{t,r}$ with volatility $\sigma_{\beta_{t}}$ evolving in continuous time $t$ and apply the series approximation as follows:
\begin{equation}
\begin{aligned}
\beta_{t,r} &= \sigma_{\beta_{t}}\sum_{i=1}^{n} Z_i\int_{0}^t\phi_i(u)du \\
&=  \sigma_{\beta_{t}} \sum_{i=1}^n Z_i( 2/T)^{1/2}\sin\{(2i-1)\pi t/2T\} \approx W_t,
\end{aligned}
\end{equation}
where the second equality follows from choosing $\phi_i$ as given in Eq~(\ref{eq: KL}) and carrying out the integration. We can then discretise this approximation using the same time-step of $\delta t$ that is used for the compartmental dynamics to obtain the following path-wise (discrete time) approximation:
\begin{equation}\label{eq: brown daily}
    \beta_{t_k,r} = \sigma_{\beta_{t}} \sum_{i=1}^n Z_{i,r}( 2/T)^{1/2}\sin\{(2i-1)\pi t_k/2T\},
\end{equation}
where $T$ is the number of days between lockdown and analysis date. 
Note that in this formulation the problem of sampling a large vector of increments $\Delta \bv{\beta}_r$ is reduced to that of sampling a $n$-dimensional vector of the coefficients $\Zb_r=(Z_{1,r},\ldots,Z_{n,r})$. From the comparison of the time $T$ marginal distributions of the true and approximate Brownian motion, as for the OU process (see Fig~\ref{Figure:wv compare}), we found $n=10$ to produce a good path-wise approximation. Thus, we used $n=10$ for the subsequent comparative evaluations. The regional parameter vector, $\thb_r = (\psi_r,I_{0_{r}},\Zb_r)$, now contains the expansion coefficients instead of the random-walk increments $\Delta \bv{\beta}_r$.

\subsection{Results: comparative evaluations}\label{sec:results covid}

We ran the MwG algorithm to target the posterior distribution in Eq~(\ref{eq:posterior}) while using the random-walk based piecewise constant transmission-potential in Eq~(\ref{eq: random ealk beta}) and the Brownian motion approximation (BMA) in Eq~(\ref{eq: brown daily}). In both cases we ran $3 \times 10^6$ iterations, discarded the first half of the iterations as burn-in and subsequently thinned the remaining samples to obtain $1000$ samples. We implemented the epidemic model in \texttt{C++}. The MwG algorithm was implemented using \texttt{Python}.

\begin{figure}[!ht]
\includegraphics[width=\textwidth,keepaspectratio=true]{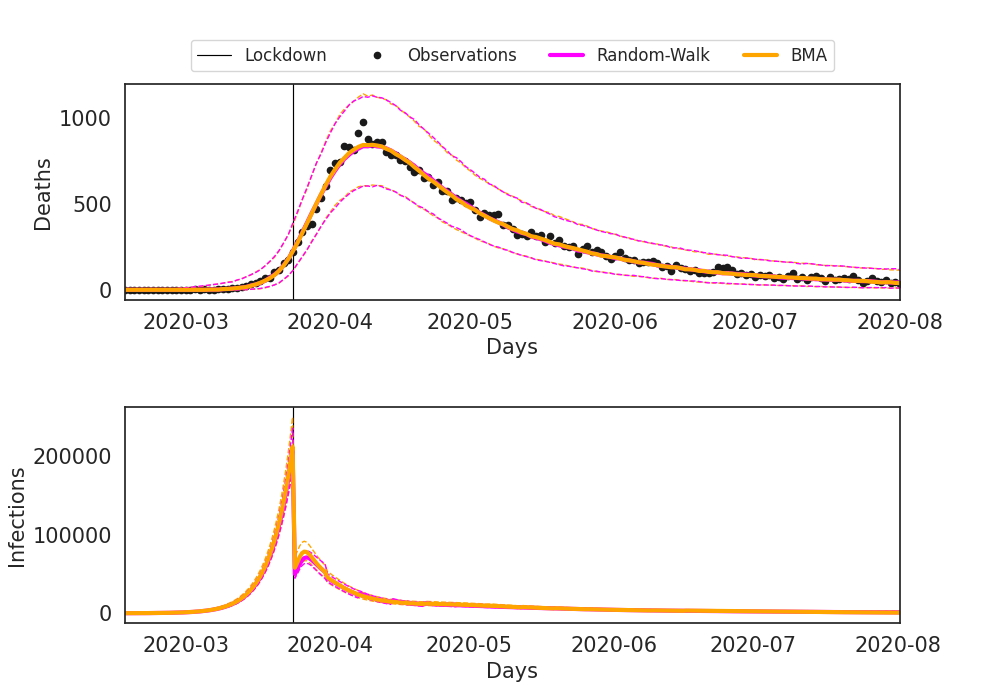}
\caption{Goodness-of-fit of daily death data (a) and the inferred latent infections (b), produced using the random-walk (magenta lines) and BMA (orange lines). These densities are summarised by the mean (solid lines) and $95\%$ credible intervals (broken lines). The black line indicates the day of lockdown in England \myrd{23} March, 2020.}
\label{coronappc compare}
\end{figure}

 Fig~\ref{coronappc compare} (a) compares, for the two alternative choices of modelling the transmission-potential, the posterior predictive distributions of the death data aggregated across all ages and regions with the observed data (see Appendix G in for region-specific plots). Clearly the goodness-of-fit is indistinguishable between the two models. In Fig~\ref{coronappc compare} (b) we show summaries of the posterior distributions of the latent infections $p(\bv{ \nni}|\bv{y^d},\bv{y^s})$, aggregated across all ages and regions (region-wise infections are shown in Appendix G) again showing close consistency across models, with the exception of a few days immediately following the lockdown where the number of infections estimated by the BMA is slightly higher.
 
Following \cite{birrell2021real}, we also obtain estimates of the effective region-specific reproduction number $R_{t_k,r}$, their weighted average $R_{t,E}$ representing the reproduction number for all of England, (formulae for these are given in Appendix D in S1 text). In Fig~\ref{rtppc compare} we show the posterior distributions for $p(R_{t,E}|\bv{y^d},\bv{y^s})$, using the two alternative models. It is evident that the estimate obtained from the BMA appears to be smoother than what is obtained using the piecewise constant model, more realistically reflecting the actual transmission process that happens in continuous time. In Table~\ref{Table2} we present infection-fatality ratio estimates from the two models, again showing close agreement across models. 
\begin{figure}[!ht]
\includegraphics[width=\textwidth,keepaspectratio=true]{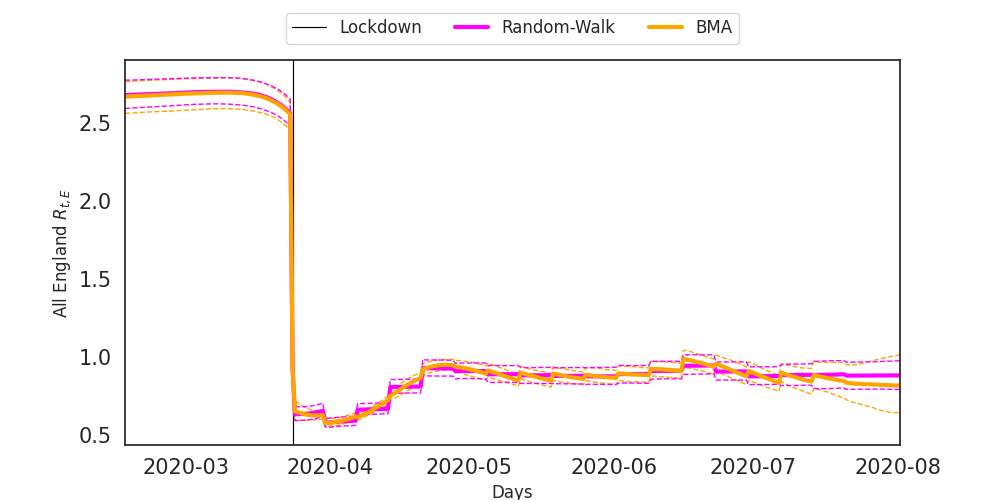}
\caption{Posterior mean (solid lines) and $95\%$ credible intervals (broken lines) for the all England reproduction number $R_{t,E}$.}
\label{rtppc compare}
\end{figure}

\begin{table}[!h]
\centering
\caption{\bf Posterior mean and $95\%$ credible intervals for the age-specific infection-fatality ratio from the random-walk and BMA models of transmission-potential. }\label{Table2}
\begin{tabular}{|c|c|c|}
\hline
    \textbf{Age group (yrs)} & \textbf{Random-walk}  & \textbf{BMA} \\
    \hline
    \multirow{2}{1.3cm}{$<5$} &0.0009\% & 0.0009\%\\
   &(0.0002\%--0.0022\%) & (0.00007\%--0.0019\%)\\
       \hline
    \multirow{2}{1.3cm}{5--14}& 0.0014\% & 0.0014\%\\
   &(0.0008\%--0.0022\%) & (0.0006\%--0.0022\%)\\
       \hline
    \multirow{2}{1.3cm}{15--24}& 0.0046\% & 0.0044\%\\
    &(0.0032\%--0.0062\%) & (0.0029\%--0.0060\%)\\
        \hline
    \multirow{2}{1.3cm}{25--44}& 0.0311\% & 0.0299\%\\
    &(0.0281\%--0.0345\%) & (0.0257\%--0.0341\%)\\
        \hline
    \multirow{2}{1.3cm}{45--64}& 0.4653\% & 0.4488\%\\
   & (0.4412\%--0.4901\%) & (0.4001\%--0.4976\%)\\
       \hline
    \multirow{2}{1.3cm}{65--74}& 3.0992\% & 2.9831\%\\
   & (2.9576\%--3.2600\%) & (2.6609\%--3.3052\%)\\
       \hline
    \multirow{2}{1.3cm}{$>74$}& 17.8161\% & 17.1086\%\\
    &(16.9632\%--18.6098\%) & (15.3604\%--18.8568\%)\\ 
    \hline
\end{tabular}
\end{table}

\paragraph{Computational gains:}

The MwG algorithm took around $78$ hours to finish for both models of the transmission-potential. However, the BMA allows inference at a faster time-scale producing a smoother estimate of $R_{t,E}$ avoiding artificial model  assumptions. Such an inference would be computationally infeasible if using a random-walk model at the more granular time-scale as in Eq~(\ref{eq: random ealk beta fast}), given the poor scaling of the RWMH proposal in high dimensions. Thus, using the series approximation we were able to extract more information about the transmission-potential and reproduction-ratio in comparison to the piecewise constant model, while incurring the same computational expense.  

Had we used the random-walk model in Eq~(\ref{eq: random ealk beta fast}), we would have had to further partition each of the regional parameter block in separate chunks to accommodate a large vector of increments $\Delta\beta_{t_{k},r}=\log\left(\beta_{t_{k},r}\right)-\log\left(\beta_{t_{k-1},r}\right)$. Consequently, multiple Gibbs moves would have been necessary to update all the increments for a randomly chosen region. This, in turn, would have increased the number of likelihood computations, involving the computationally expensive updates of the transmission model, exponentially at each MCMC iteration. 

\section{Discussion}\label{sec:Discussion}

By modelling the force of infection as the function of a time-varying transmission-potential we can incorporate extrinsic, un-modelled effects in the description of the transmission process within a compartmental model. Describing this transmission-potential, in turn, as a stochastic process, a diffusion in particular, we can inject environmental stochasticity in an otherwise deterministic model. In this paper we proposed a path-wise approximation of a diffusion process as an alternative to modelling the dynamics of the transmission-potential as a SDE. Through the path-wise approximation we arrive at a random ODE approximating the SDE. As a function of its parameters, the path (solution) of an ODE is completely deterministic. As a result inference of the transmission-potential is simplified, with no need to solve a missing data problem using a computationally expensive data-augmentation procedure.

We demonstrate the efficacy of the proposed path-wise approximation using two epidemic models. In the first one, an influenza model, we replaced an OU SDE with an equivalent path-wise approximation. We noticed similar inference outcomes in terms of parameter estimates and goodness-of-fit using the SDE and its ODE approximation. However, for the latter we observed orders-of-magnitude improvement in computational efficiency. 

We then applied the path-wise approximation to replace a Gaussian random-walk with a discretised path-wise approximation of Brownian motion to model the transmission-potential within a compartmental model of COVID-19 pandemic spread in England. Again we noticed consistent estimates of crucial unknown quantities such as infection-fatality rate, latent infections and a time-varying estimate of the reproduction number. In addition, the path-wise approximation allows the transmission-potential to be modelled at a more granular time-scale providing a smooth estimate of the effective reproduction number. This would be impossible to achieve using the random-walk model due to an exorbitant computational burden.

As an alternative to using our path-wise approximation of Brownian motion to model the transmission-potential, at a faster time-scale, we could have used a different MCMC algorithm, such as the No-U-Turn sampler \citep{hoffman2014no}, that is known to perform well for high dimensional problems. This algorithm proposes a move based on the gradient of the target density. Evaluating gradients, however, for the COVID-19 model is challenging as this requires, in addition to extra computations, a complete re-implementation of the model using an automatic differentiation package. However, for modelling studies where such re-implementation is straightforward, we like to point out that by applying a path-wise approximation of a diffusion process we are left with the task of sampling from a posterior distribution with a standard Gaussian prior (over the coefficients). The No-U-Turn sampler generally excels at this task.

In this paper we have used simple diffusion models whose transition densities are known analytically. However, if additional prior information about the force of infection is available, then such information can be incorporated in more complex nonlinear SDEs as models of the time-varying transmission-potential. Our methodology can be seamlessly applied in such cases to arrive at a path-wise approximation of such complex diffusion processes.

\section{Software}
\label{sec5}

Code and data supporting the experiment with the SIR model in Section \ref{sec:Evaluation}, and the code for running the COVID-19 model in Section  \ref{sec:Application: modelling COVID-19 outbreak in England} is available at \url{https://github.com/sg5g10/envstoch}. Requests to access the non-publicly available data used for the COVID-19 model in section 5, are handled by the UKHSA Office for Data Release (ODR) \url{https://www.gov.uk/government/publications/accessing-ukhsa-protected-data}



\bibliographystyle{apalike}
\bibliography{arxiv}
\section*{Appendix A: Fourier expansion of Brownian motion}\label{sec: Fourier expansion of Brownian motion}
By the definition of an It\^o integral, within a time interval $[0,T]$ a standard Brownian motion can be written as \citep{luo2006wiener,LyonsSS12}:
\begin{equation}\label{eq: Brownian indicator}
    W_t = \int_{0}^t dW_s = \int_{0}^T \mathbb{I}_{[0,t]}(s)dW_s,
\end{equation}
where $\mathbb{I}_{[0,t]}(\cdot)$ is the indicator function. Suppose $\{\phi_i\}_{i=1}^{\infty}$ is a complete orthonormal basis of $L^2[0,T]$. 
We can interpret $\mathbb{I}_{[0,t]}$ as an element of $L^2[0,T]$, and expand it in terms of the basis functions:
\begin{equation}\label{eq:brown innerp}
\begin{aligned}
   \mathbb{I}_{[0,t]}(s)&=\sum_{i=1}^{\infty} \left\langle \mathbb{I}_{[0,t]}(\cdot),\phi_i(\cdot)\right\rangle\phi_i (s)\\ &=\sum_{i=1}^{\infty} \Big(\int_{0}^t\phi_i(u)du\Big)\phi_i (s).
 \end{aligned}
\end{equation}
Substituting (\ref{eq:brown innerp}) into (\ref{eq: Brownian indicator}) we see that:
\begin{equation}\label{eq: Brownian series}
    W_t = \sum_{i=1}^\infty\Big(\int_{0}^T\phi_i(s)dW_s\Big)\int_{0}^t\phi_i(u)du.
\end{equation}

\section*{Appendix B: Adaptive MCMC}\label{sec:appendix a}

In an adaptive MCMC algorithm optimal values of the proposal density is learnt on the fly using past samples from the Markov chain. Different mechanisms can be used to adapt or learn the parameters of the proposal. \citep{andrieu2008tutorial} proposed a general framework for constructing adaptive MCMC algorithms that rely on the \textit{stochastic approximation} method \citep{robbins1951stochastic} for learning the proposal's parameters on the fly.

Consider in general the proposal density $q_{\phi}(\thb^{j+1}|\thb^j)$ parameterised by $\phi$. Let us also define a suitable objective function 
\begin{equation}
h(\phi):= \mathbb{E}^{\phi}\big[H(\phi,\thb^0,\thb^1,\ldots,\thb^{j},\thb^{j+1})\big], 
\end{equation}
 that expresses some measure of the statistical performance of the Markov chain in its stationary regime. The expectation is with respect to a $\phi$ dependent distribution. For example, the coerced acceptance probability is often used as the objective:
\begin{equation}\label{eq: coerced}
H(\phi,\thb^0,\thb^1,\ldots,\thb^{j},\thb^{j+1})=\underbrace{\mathrm{min}\left\{1,\frac{\pi(\thb^{j+1})}{\pi(\thb^{j})}\frac{q_{\phi}(\thb^{j}|\thb^{j+1})}{q{\phi}(\thb^{j+1}|\thb^{j})}\right\}}_{=:\alpha^j}-\bar{\alpha},
\end{equation}
where $\pi(\thb)$ is the target distribution and $\bar{\alpha}$ is the approximate optimal expected acceptance probability in the stationary regime. For the Gaussian proposal $q:= \mathcal{N}(\thb^{j+1}|\thb^{j},\bv{\Sigma}^{j})$, with its parameter $\phi$ being the covariance $\bv{\Sigma}^{j}$, the following objective function:
\begin{equation}\label{eq: moment matching}
  H(\bv{\Sigma}^{j},\thb^{j+1}) = \thb^{j+1}\thb^{{j+1}^{'}} - \bv{\Sigma}^{j},   
\end{equation}
corresponds to matching the moments of the proposal with that of the target. Here by $a^{'}$ we denote the transpose of the vector $a$.
 
 Optimal exploration of $\pi(\thb)$ can thus be formulated as finding the root $\bar{\phi}$ of the following equation: $h(\phi)=0$. The challenge here is to devise an algorithm to find the roots of $h(\phi)$, which involves both integration and optimisation. \cite{andrieu2008tutorial} suggested using the stochastic approximation method \citep{robbins1951stochastic} which is tailored to this situation:
\begin{equation}
\begin{aligned}
    \phi^{j+1} &=  \phi^{j} + \delta^j H(\phi^j,\thb^0,\thb^1,\ldots,\thb^{j},\thb^{j+1})\\
    &= \phi^{j} + \delta^{j+1}h(\phi) + \delta^{j+1} H(\phi^j,\thb^0,\thb^1,\ldots,\thb^{j},\thb^{j+1}) - \delta^{j+1}h(\phi)\\
                &= \phi^{j} + \delta^{j+1}h(\phi) + \delta^{j+1}\xi^{j+1},
\end{aligned}
\end{equation}
where $\xi^{j+1}:=\big[H(\phi^j,\thb^0,\thb^1,\ldots,\thb^{j},\thb^{j+1})-h(\phi)\big]$ is usually referred to as the \textit{noise term} and $\delta^j$ is a decreasing sequence (a step-size parameter). If the noise term $\xi^{j+1}$ averages to zero as $j\rightarrow \infty$, the above recursion will converge to the root $\bar{\phi}$ (or at least oscillate around it) when the following conditions hold:
\begin{equation}
    \sum_{j=0}^{\infty} \delta^j = \infty \quad \operatorname{and} \quad \sum_{j=0}^{\infty} (\delta^j)^2 < \infty.
\end{equation}
Combining the above objective functions and using the stochastic approximation we have the following recursions for adapting a random-walk proposal with a global scaling $\lambda^{j}$, $ \mathcal{N}(\thb^{j+1}|\thb^{j},\lambda^{j}\bv{\Sigma}^{j})$, as \citep{andrieu2008tutorial}:
\begin{equation}\label{eq: adaptive MCMC}
\begin{aligned}
\log(\lambda^{j+1}) &= \log(\lambda^{j}) + \delta^{j+1}(\alpha^{j+1} - \bar{\alpha}) \\
\bv{\mu}^{j+1} &= \bv{\mu}^{j} + \delta^{j+1}(\thb^{j+1} - \mu^{j}) \\
\bv{\Sigma}^{j+1} &= \bv{\Sigma}^{j} + \delta^{j+1}(\thb^{j+1}\thb^{{j+1}^{'}}  - \bv{\Sigma}^{j}),
\end{aligned}
\end{equation}
where the recursion in the first equation, trying to adapt the global scaling, is based on the coerced accepted probability objective in \eqref{eq: coerced} and the following two equations are minimising the moment matching objective in \eqref{eq: moment matching}.  

By choosing a decreasing sequence $\{ \delta^{j}\}_{j=0}^\infty$ of step-sizes it is ensured that the adaptation declines over time, also known as \textit{vanishing adaptation} \citep{andrieu2008tutorial}, and the Markov chain converges to the correct stationary distribution. For all the experiments we have consistently used the following schedule:
\begin{equation}
    \delta^{j}= j^{-0.6},
\end{equation}
which was shown to work particularly well for nonlinear differential equation models in \cite{johnstone2016uncertainty}.

\section*{Appendix C: Simulation study for influenza epidemic}
\begin{figure}[!ht]
\includegraphics[width=\textwidth,keepaspectratio=true]{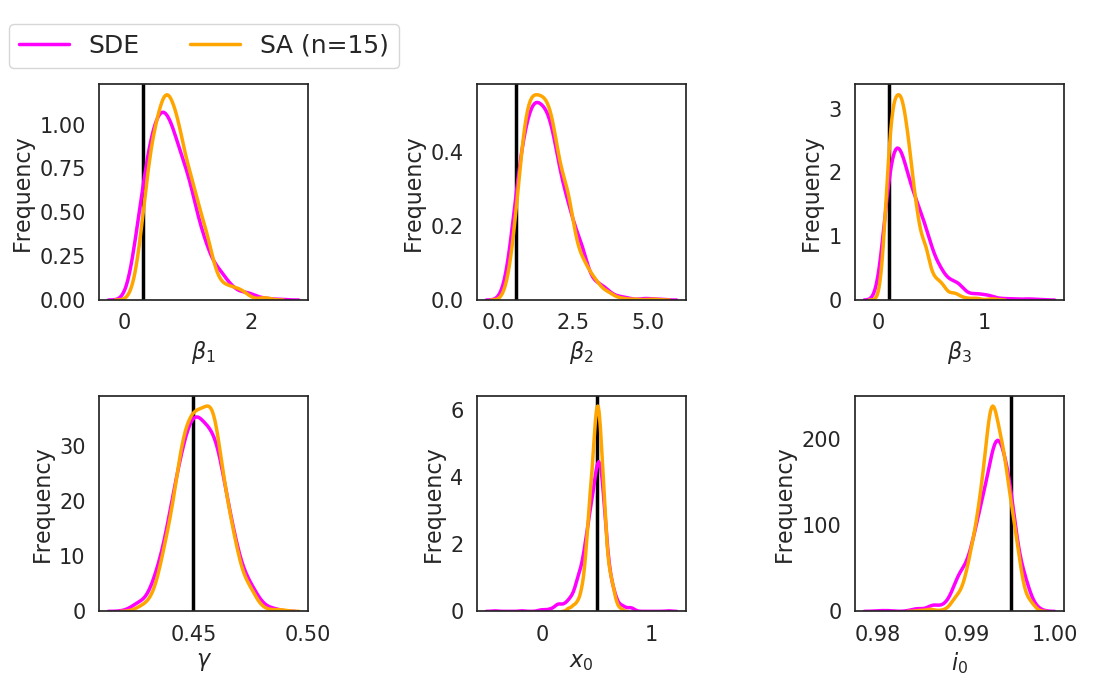}
\caption{\textbf{Simulated dataset 1}: Posterior marginal densities of the parameters obtained using the \textbf{SDE} and the \textbf{SA} (with $n=15$ basis function). These densities are summarised using a kernel density estimate. The black line in each of the plots demarcate the generative parameter value.}
\label{oumarg compare}
\end{figure}

\begin{figure}[!ht]
\includegraphics[width=\textwidth,keepaspectratio=true]{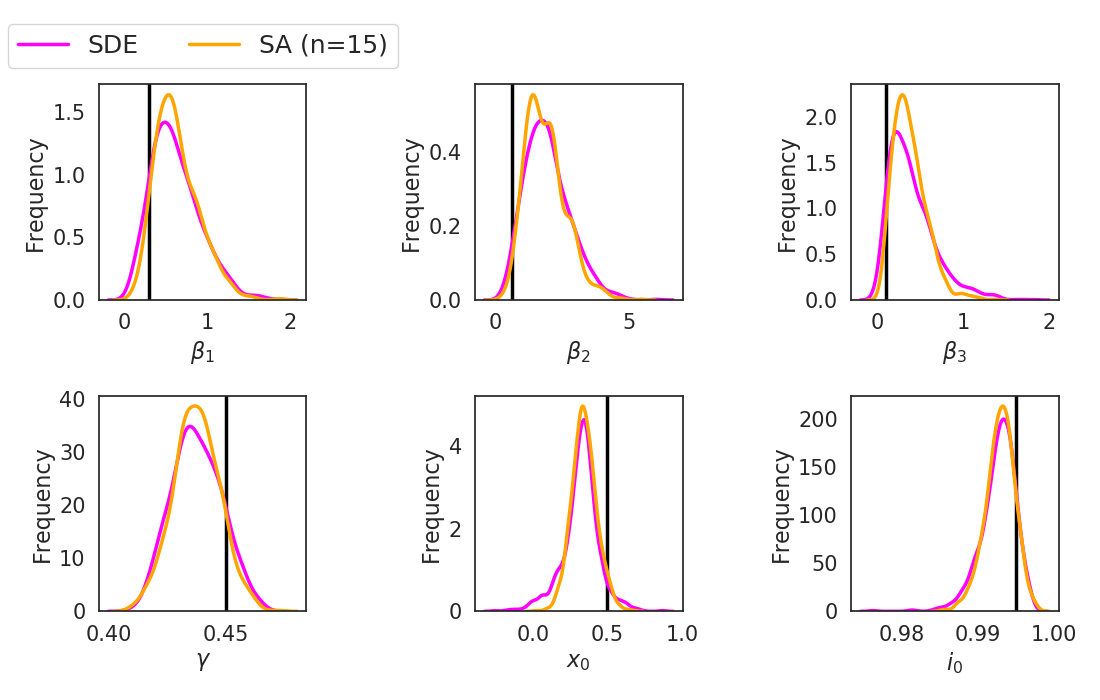}
\caption{\textbf{Simulated dataset 2}: Posterior marginal densities.}
\label{oumarg2 compare}
\end{figure}

\begin{figure}[!ht]
\includegraphics[width=\textwidth,keepaspectratio=true]{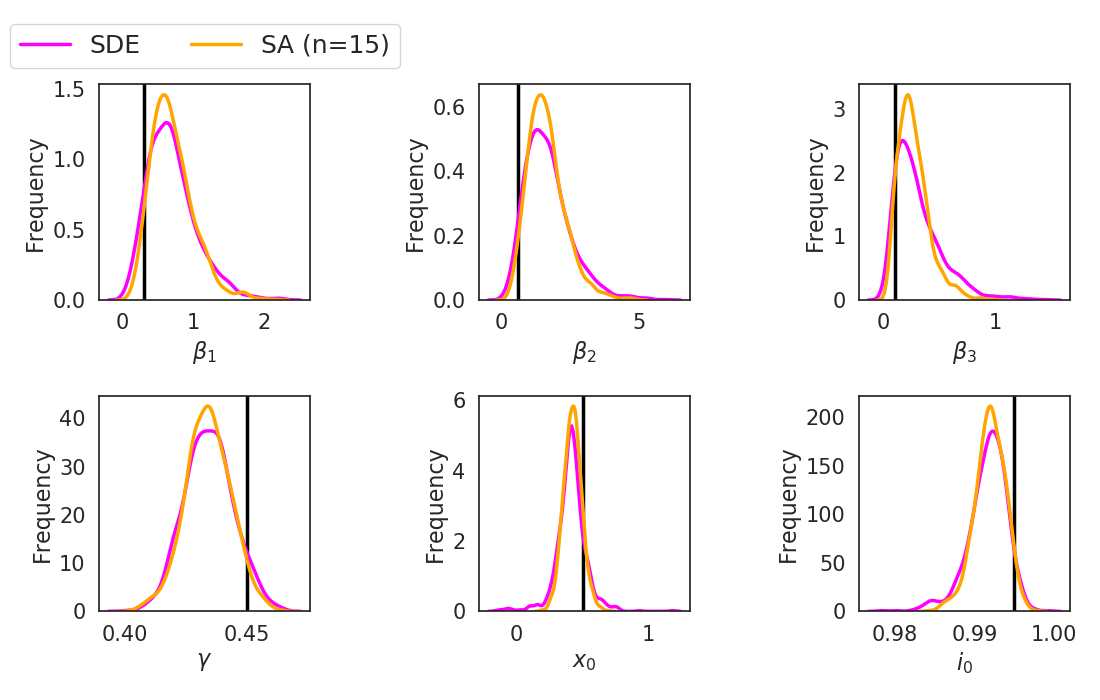}
\caption{\textbf{Simulated dataset 3}: Posterior marginal densities.}
\label{oumarg3 compare}
\end{figure}

Using a real dataset we are oblivious to the ground truth of the estimated quantities. Thus, we have also carried out a detailed simulation study where we have used simulated datasets that mimic the influenza epidemic used in the main text. We generated three simulated epidemics using the model in Equation (2.2), in the main text, on the same time period $T=14$ days, and used the same population size $N=763$, as the real influenza epidemic. We chose parameter values that generate an epidemic curve similar to the real dataset. These generative parameter values are shown in Figure \ref{oumarg compare}--\ref{oumarg3 compare}. We then proceed to fit the two alternative models using the inferential setup discussed in the main text.


In Figure \ref{oumarg compare}--\ref{oumarg3 compare} we compare the marginal densities of the parameters obtained using the \textbf{SDE} and \textbf{SA} counterparts, for each of the simulated datasets. Clearly the estimates match well and generative parameter values are recovered.

\begin{figure}[!ht]
\includegraphics[width=\textwidth,keepaspectratio=true]{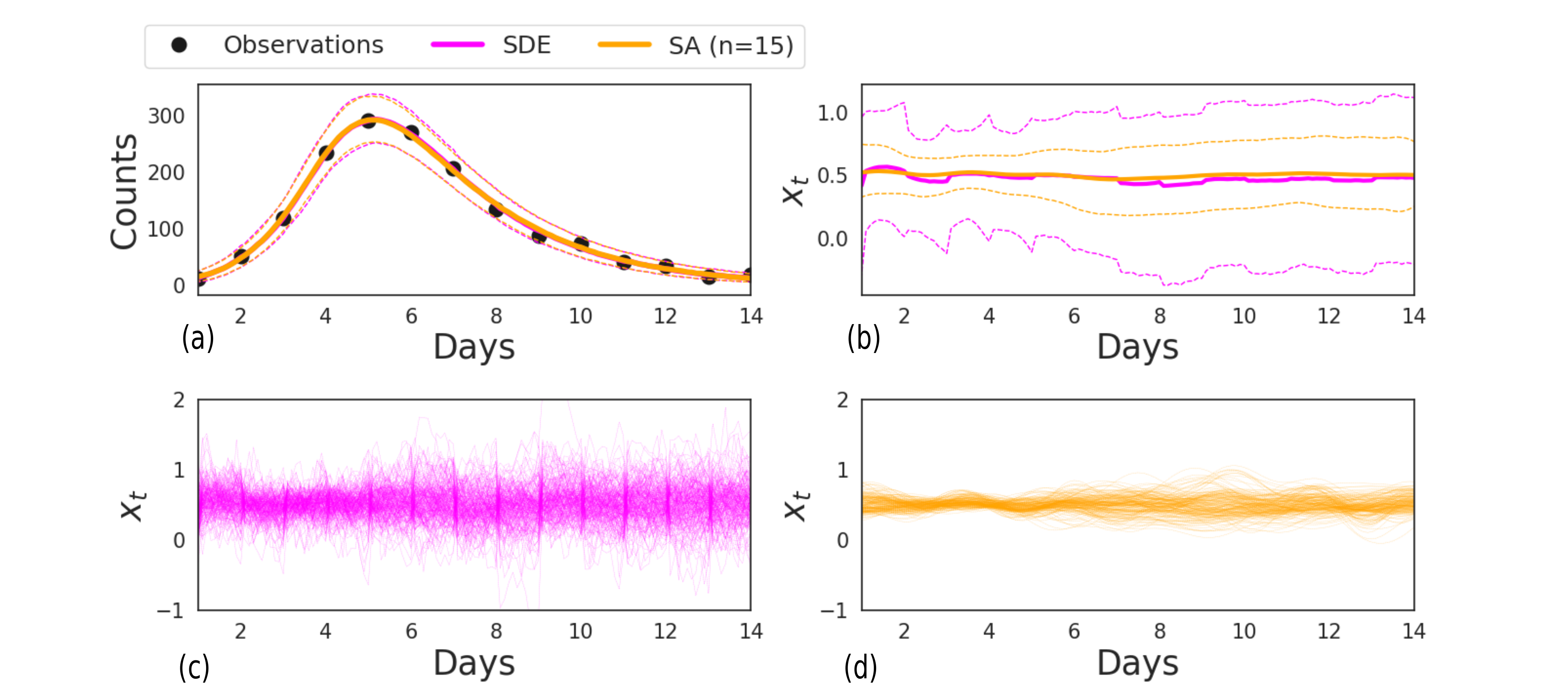}
\caption{\textbf{Simulated dataset 1}: Goodness-of-fit (a); posterior distribution of the latent diffusion paths corresponding to the \textbf{SDE} and \textbf{SA} counterparts (b), with densities summarised by the mean (solid lines) and $95\%$ credible intervals (broken lines); and samples from the posterior distribution of the latent diffusion paths, \textbf{SDE} (c) and \textbf{SA} (d)}
\label{ouppc compare}
\end{figure}
\begin{figure}[!ht]
\includegraphics[width=\textwidth,keepaspectratio=true]{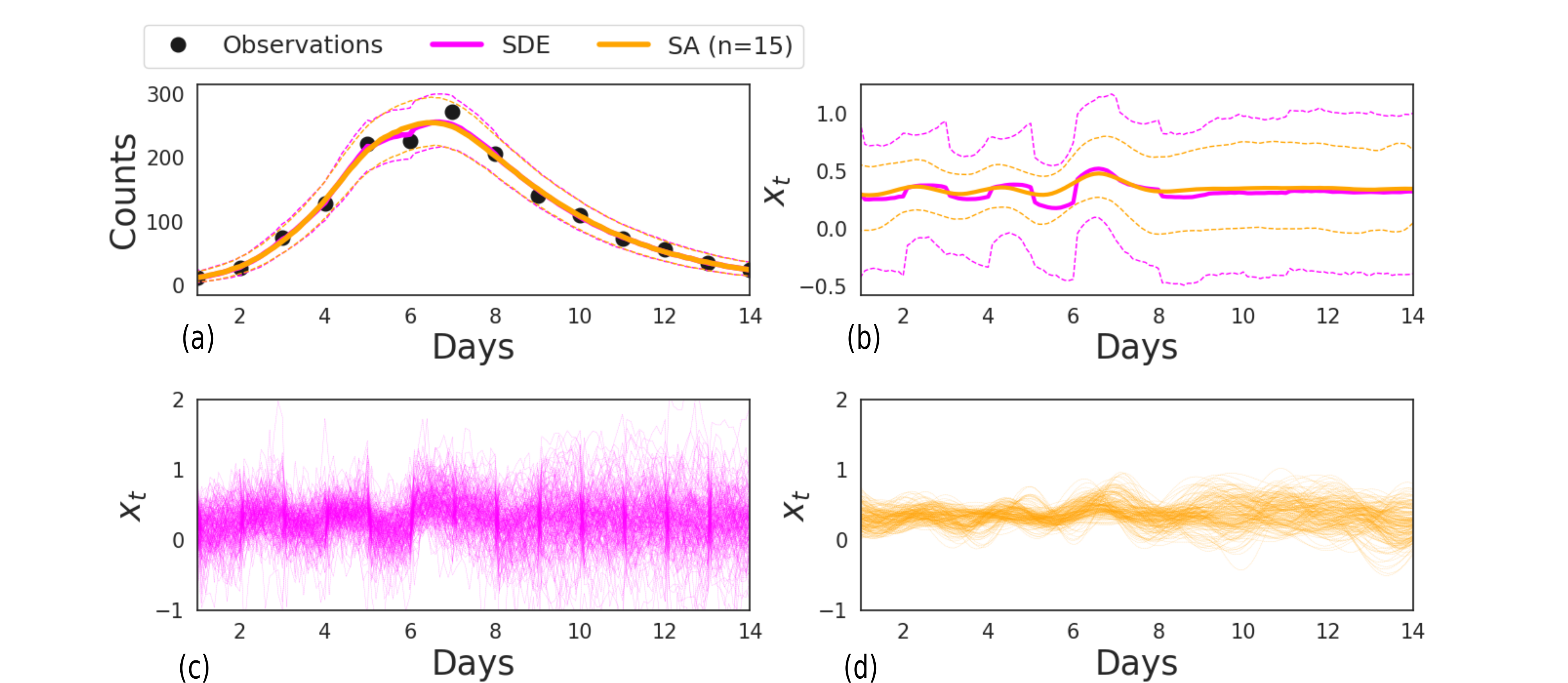}
\caption{\textbf{Simulated dataset 2}: Comparison of the goodness-of-fit}
\label{ouppc2 compare}
\end{figure}
\begin{figure}[!ht]
\includegraphics[width=\textwidth,keepaspectratio=true]{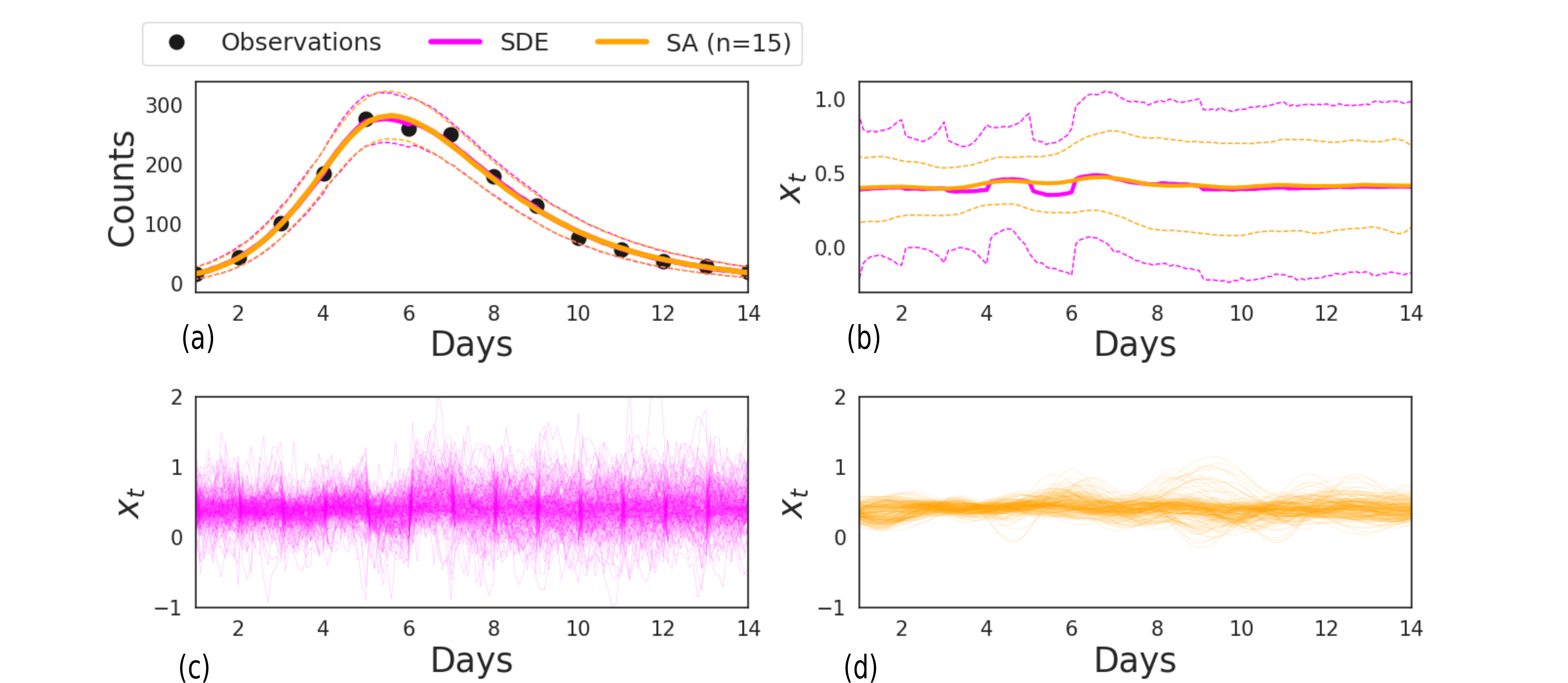}
\caption{\textbf{Simulated dataset 3}: Comparison of the goodness-of-fit}
\label{ouppc3 compare}
\end{figure}

Furthermore, in Figure \ref{ouppc compare}--\ref{ouppc3 compare} we compare the goodness-of-fit. As was found for the real dataset, we observe little disagreement between the epidemic curves obtained using the \textbf{SDE} and the \textbf{SA}, but for the posterior distribution of the latent diffusion paths we noticed, for all the datasets, that the credible intervals are narrower for the \textbf{SA}. For all these datasets, the posterior means, and the draws of the sample path, of the two models match well.

\section*{Appendix D: Calculating a time-varying reproduction number}

The estimate of the contact-rate $\beta_{t_k,r}$ is used to derive an estimate of a time-varying reproduction number. Firstly, using the formula of \cite{wearing2005appropriate}, the initial reproduction number $R_{0,r}$ is estimated as follows:
\begin{equation}
    R_{0,r}= \psi_r d_I \frac{\left (\frac{\psi_r d_L}{2} + 1 \right)^2}{1 - \frac{1}{\left ( \frac{\psi_r d_I}{2} + 1 \right )^2}}.
\end{equation}
Over time the value of the reproduction number will change as contact patterns shift and the supply of susceptible individuals deplete. The time-$t$ reproduction number is then estimated using the following formula:
\begin{equation}\label{eq: rt formula}
  R_{t_k,r} = \begin{cases} R_{0,r} \frac{R^*_{t_k,r}}{R^*_{0,r}} & \text{if $t_k < \tlock$}\\
    \beta_{t_k,r}R_{0,r} \frac{R^*_{t_k,r}}{R^*_{0,r}} & \text{if $t_k \geq \tlock$}
    \end{cases}
\end{equation}
where $\tlock$ indicates the time-point corresponding to the lockdown. $R^*_{t_k,r}$ is the dominant eigenvalue of the time $t_k$ next-generation matrix, $\Lambda_{k,r}$, with elements:
    \begin{equation}
      \left(\Lambda_{k,r}\right)_{ij} = S_{r,{t_k},i}  \vec{C}^{t_k}_{r,ij} d_I,
    \end{equation}
where $\vec{C}^{t_k}_{r,ij}$ is a region-specific time-varying contact matrix, see \cite{birrell2021real} for further details on these matrices.

To get an `all England' value for $R_{t_k,E}$ a weighted average of the regional $R_{t_k,r}$ is calculated, where the weights are given by the sum of the infections in each region:
\begin{equation}\label{eq: all england rt}
R_{t_k,E} = \frac{\sum_r R_{t_k, r} \sum_i \nni_{r,t_k,i}}{\sum_r \sum_i \nni_{r,t_k,i}}.
\end{equation}

\section*{Appendix E: Priors for the COVID-19 model}

The priors for the global and regional parameters for the COVID-19 model are listed in Table \ref{tbl:pars}. We used the same priors as was used in \cite{birrell2021real}. Note that we also used the same prior for the volatility of both the piecewise constant random-walk and the Brownian motion model of the transmission-potential. 

\begin{table}[!ht]
\caption{Model parameters with assumed prior distributions or fixed values, as was used in \cite{birrell2021real}.
\label{tbl:pars}}
{\tabcolsep=4.25pt
\begin{tabular}{p{7.3cm}p{7.3cm}}
Name&Prior source\\
\toprule
    Over-dispersion, $\eta$        &Uninformative $\operatorname{Gamma}(1, 0.2)$.\\
    Mean infectious period, $d_I$     &2 + $\operatorname{Gamma}(1.43, 0.549)$.\\
    Infection-fatality rate for age $<5$: $p_1$   & $\operatorname{Beta}(1,62110.8012)$.\\
    Infection-fatality rate for age, $5-14$: $p_2$   &$\operatorname{Beta}(1,23363.4859)$.\\
    Infection-fatality rate for age $15-24$: $p_3$   &$\operatorname{Beta}(1,5290.0052)$.\\
    Infection-fatality rate for age $25-44$: $p_4$   &$\operatorname{Beta}(1,1107.6474)$.\\
    Infection-fatality rate for age $45-64$: $p_5$   &$\operatorname{Beta}(1,120.9512)$.\\
    Infection-fatality rate for age $65-74$: $p_6$   &$\operatorname{Beta}(1,31.1543)$.\\
    Infection-fatality rate for age $>74$: $p_7$   &$\operatorname{Beta}(9.5,112)$.\\
    Serological test sensitivity, $\ksens$& $\operatorname{Beta}(71.5,29.5)$.\\
    Serological test specificity, $\kspec$& $\operatorname{Beta}(777.5,9.5)$.\\
    Exponential growth, $\psi_r$ & $\operatorname{Gamma}(31.36, 224)$.\\
    Log of initial infectives, $\log{I_{0,r}}$ & $\mathcal{N}(-17.5, 1.25^2)$.\\
    Volatility of transmission-potential, $\sigma_{\beta_{w}},\sigma_{\beta_{t}}$&$\operatorname{Gamma}(1, 100)$.\\
    \hline
    Mean latent period, $d_L$&3 days (fixed not estimated).\\
\bottomrule
\end{tabular}}
\end{table}

\section*{Appendix F: Pseudocode of the MwG algorithm}

The pseudocode listed in Algorithm \ref{alg:MwG} describes the Metropolis-within-Gibbs algorithm for sampling from the posterior distribution $p(\thb_g,\thb_1,\ldots,\thb_{n_{r}}|\bv{y^d},\bv{y^s})$ of the global $\thb_g$ and regional $\thb_1,\ldots,\thb_{n_{r}}$ parameters of the COVID-19 model. For each parameter group $\thb_g,\thb_1,\ldots,\thb_{n_{r}}$ we use a proposal with a different set of parameters that are adapted through the mechanism described in \eqref{eq: adaptive MCMC}.

\begin{algorithm}
\small
   \caption{A random-scan adaptive Metropolis-within-Gibbs sampler}
   \label{alg:MwG}
\begin{algorithmic}
   \STATE {\bfseries Input:} Number of iterations $J$; data $\bv{y^d},\bv{y^s}$; optimal acceptance rate $\bar{\alpha}$.
 \STATE Initialise the regional $\thb^{0}_1,\ldots,\thb^{0}_{n_{r}}$ and global parameters $\thb^{0}_g$. 
 \STATE Initialise the regional proposal parameters $\lambda^{0}_1,\ldots,\lambda^{0}_{n_{r}}$, $\bv{\mu}^{0}_1,\ldots,\bv{\mu}^{0}_{n_{r}}$ and $\bv{\Sigma}^{0}_1,\ldots,\bv{\Sigma}^{0}_{n_{r}}$.
 \STATE Initialise the global proposal's parameters $\lambda^{0}_g$, $\bv{\mu}^{0}_g$ and $\bv{\Sigma}^{0}_g$.
   \FOR{$j=0$ {\bfseries to} $J-1$}
   \STATE \textbf{Global move}: 
   \begin{enumerate}
       \item Draw $\thb^*_g \sim \mathcal{N}(\thb^j_g,\lambda^{j}_g\bv{\Sigma}^j_g)$ and set $\thb^{j+1}_g=\thb^*_g$ with probability $\alpha^j_g=\operatorname{min}\Big\{1,\frac{p(\thb^*_g|\bv{y^d},\bv{y^s})}{p(\thb_g|\bv{y^d},\bv{y^s})}\Big\}$, otherwise $\thb^{j+1}_g=\thb^j_g$.
   \end{enumerate}
   \STATE \textbf{Regional move}: \begin{enumerate}
       \item Draw $r^{*}\sim \operatorname{Uniform}(1,n_r)$.
       \item Draw $\thb^*_{r^{*}}\sim \mathcal{N}(\thb^j_{r^{*}},\lambda^{j}_{r^{*}}\bv{\Sigma}^j_{r^{*}})$ 
       and set $\thb^{j+1}_{r^{*}}=\thb^*_{r^{*}}$ with probability $\alpha^j_{r^{*}}=\operatorname{min}\Big\{1,\frac{p(\thb^*_{r^{*}}|\bv{y^d},\bv{y^s})}{p(\thb^j_{r^{*}}|\bv{y^d},\bv{y^s})}\Big\}$, otherwise $\thb^{j+1}_{r^{*}}=\thb^{j}_{r^{*}}$.
       \item Set $\thb^{j+1}_{{n_{r}}\setminus{r^{*}}}= \thb^j_{{n_{r}}\setminus{r^{*}}}$, where the symbol $A\setminus a$ denotes all elements of the set $A$ except $a$.
   \end{enumerate}
    \STATE \textbf{Adaptation}: \begin{enumerate}
    \item Adapt global proposal's parameters:\\
    
    \begin{equation}
    \begin{aligned}
    \log(\lambda^{j+1}_g) &= \log(\lambda^{j}_g) + \delta^j(\alpha^j_g - \bar{\alpha}) \\
    \bv{\mu}^{j+1}_g &= \bv{\mu}^j_g + \delta^j(\thb^{j+1}_g - \bv{\mu}^j_g) \\
    \bv{\Sigma}^{j+1}_g &= \bv{\Sigma}^j_g + \delta^j(\thb^{j+1}_g\thb^{{j+1}^{'}}_g - \bv{\Sigma}^j_g).
    \end{aligned}
    \end{equation}
    \item Adapt proposal's parameters for region ${r^{*}}$:\\
    \begin{equation}
    \begin{aligned}
    \log(\lambda^{j+1}_{r^{*}}) &= \log(\lambda^{j}_{r^{*}}) + \delta^j(\alpha^j_{r^{*}} - \bar{\alpha}) \\
    \bv{\mu}^{j+1}_{r^{*}} &= \bv{\mu}^j_{r^{*}} + \delta^j(\thb^{j+1}_{r^{*}}- \bv{\mu}^j_{r^{*}}) \\
    \bv{\Sigma}^{j+1}_{r^{*}} &= \bv{\Sigma}^j_{r^{*}} + \delta^j(\thb^{j+1}_{r^{*}}\thb^{{j+1}^{'}}_{r^{*}} - \bv{\Sigma}^j_{r^{*}}).
    \end{aligned} 
    \end{equation}
    \item Set $\lambda^{j+1}_{{n_{r}}\setminus{r^{*}}}= \lambda^j_{{n_{r}}\setminus{r^{*}}}$, $\bv{\mu}^{j+1}_{{n_{r}}\setminus{r^{*}}}= \bv{\mu}^j_{{n_{r}}\setminus{r^{*}}}$ and $\bv{\Sigma}^{j+1}_{{n_{r}}\setminus{r^{*}}}= \bv{\Sigma}^j_{{n_{r}}\setminus{r^{*}}}$.
    \end{enumerate}
   \ENDFOR
   \STATE {\bfseries Output:} $\{\thb^j_g,\thb^j_1,\ldots,\thb^j_{n_{r}}\}_{j=0}^{J-1} $.
\end{algorithmic}
\end{algorithm}

\section*{Appendix G: Goodness-of-fit as per regions of England}

\begin{figure}[!p]
\includegraphics[width=\textwidth,keepaspectratio=true]{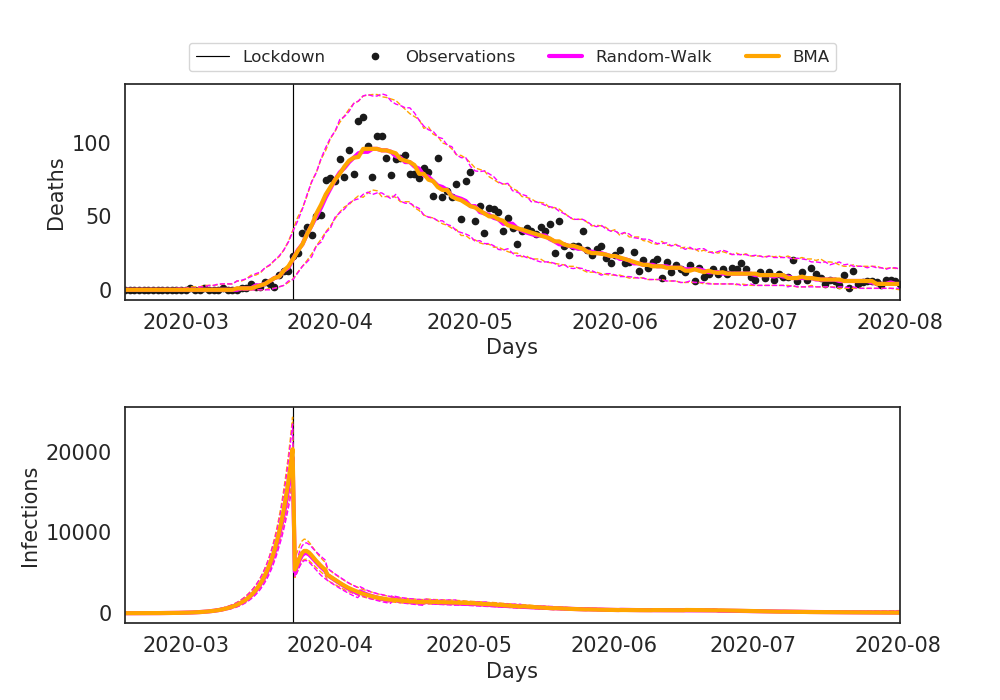}
\caption{Goodness-of-fit of daily death data (a) and the inferred latent infections (b), produced using the random-walk (magenta lines) and BMA (orange lines) for the region \textbf{East of England}. These densities are summarised by the mean (solid lines) and $95\%$ credible intervals (broken lines). The black line indicates the day of lockdown in England \myrd{23} March, 2020.}
\label{gof1}
\end{figure}

\begin{figure}[!ht]
\includegraphics[width=\textwidth,keepaspectratio=true]{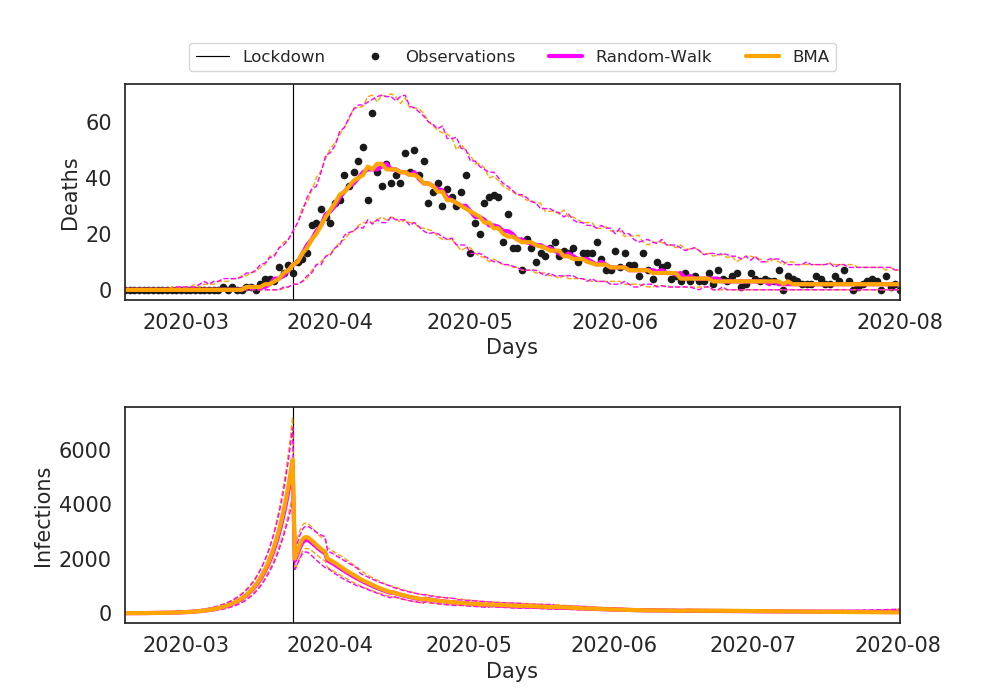}
\caption{Goodness-of-fit of daily death data (a) and the inferred latent infections (b) for the region \textbf{South West}. }
\label{gof7}
\end{figure}

In Figure \ref{gof1} -- \ref{gof7} we show the posterior predictive distributions of the number of deaths and the posterior distribution of the latent infection for each region respectively. We have aggregated the results across ages.

\section*{Appendix H: Maximum mean discrepancy}

For any given probability distribution $\mathbb{P}$ on a domain $\mathcal{X}$ its kernel embedding is defined as $\mu_{\mathbb{P}} =\mathbb{E}_{X \sim \mathbb{P}} k(\cdot,\thb)$ \citep{muandet2017kernel}, an element of reproducing kernel Hilbert space $\mathcal{H}$ associated with a positive definite kernel function $k:\mathcal{X} \times \mathcal{X} \rightarrow \mathbb{R}$. Such an embedding exists for any $\mathbb{P}$ whenever $k$ is bounded. Given two probability distributions $\mathbb{P}$  and $\mathbb{Q}$ the maximum mean discrepancy (MMD) is the Hilbert space distance between their kernel embedding $\mu_{\mathbb{P}}$ and $\mu_{\mathbb{Q}}$. Considering that we have two set of samples $\{X_i\}_{i=1}^n$ and $\{Y_i\}_{i=1}^m$ from corresponding distributions $\mathbb{P}$ and $\mathbb{Q}$ respectively, then the MMD between $\mathbb{P}$ and $\mathbb{Q}$ is given by \citep{gretton2012kernel}
\begin{equation}\label{eq:MMD}
\begin{aligned}
    MMD^2(\mathbb{P},\mathbb{Q})&=|| \mu_{\mathbb{P}} - \mu_{\mathbb{Q}}||_{\mathcal{H}}\\
    &=\frac{1}{n(n-1)}\sum_{i=1}^n \sum_{j\ne i}^m k(X_i,X_j) + \frac{1}{m(m-1)}\sum_{i=1}^n \sum_{j\ne i}^m k(Y_i,Y_j)-\frac{2}{nm}\sum_{i=1}^n \sum_{j=1}^m k(X_i,Y_j).
\end{aligned}
\end{equation}
The $MMD^2(\mathbb{P},\mathbb{Q})=0$ iff $\mathbb{P}=\mathbb{Q}$, following the properties of kernel embedding. The kernel embedding captures all the necessary information about a distribution \citep{muandet2017kernel}, thus the distance between two embedding would naturally highlight the discrepancy more efficiently in the tail regions of the distributions under comparison. In this paper we used an exponentiated quadratic kernel given by
\begin{equation}\label{eq: rbf kernel}
   k (X,X')= \exp{\Big(\frac{||X - X'||^2}{\rho^2}\Big)},
\end{equation}
where $\rho$ is a hyperparameter. We set $\rho$ to the median distance among the samples.


\begin{figure}[!ht]
\includegraphics[width=\textwidth,keepaspectratio=true]{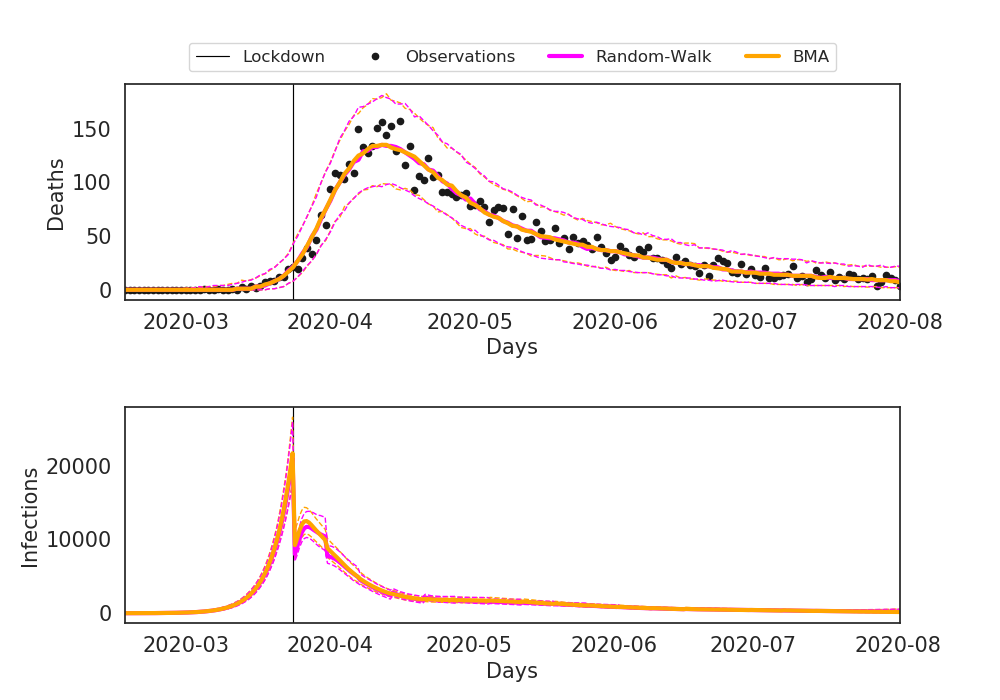}
\caption{Goodness-of-fit of daily death data (a) and the inferred latent infections (b) for the region \textbf{North West}. }
\end{figure}

\begin{figure}[!ht]
\includegraphics[width=\textwidth,keepaspectratio=true]{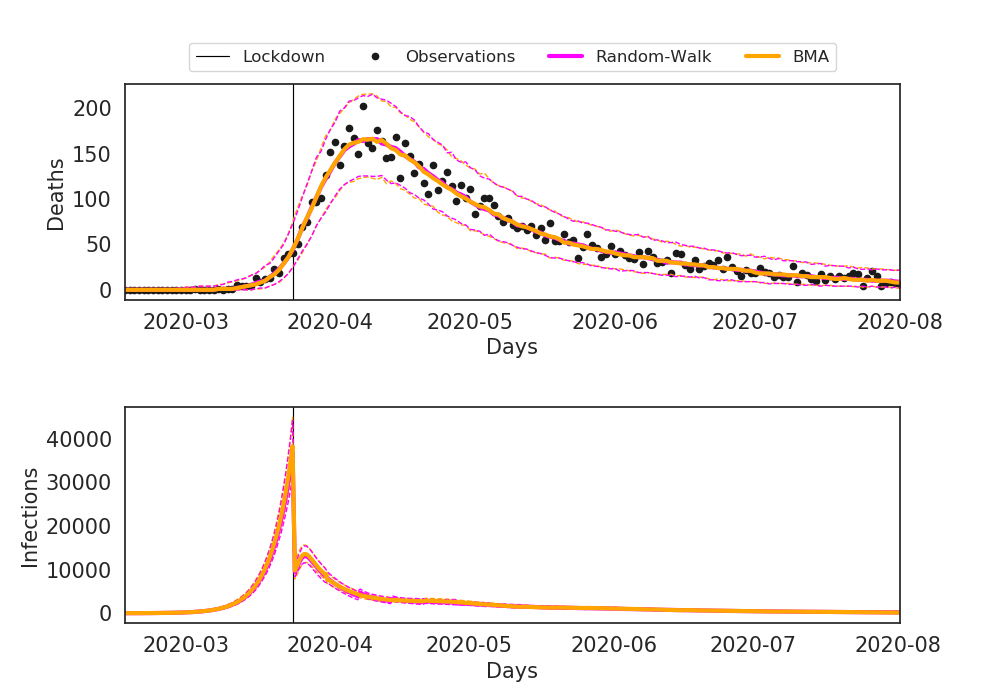}
\caption{Goodness-of-fit of daily death data (a) and the inferred latent infections (b) for the region \textbf{Midlands}.}
\end{figure}

\begin{figure}[!ht]
\includegraphics[width=\textwidth,keepaspectratio=true]{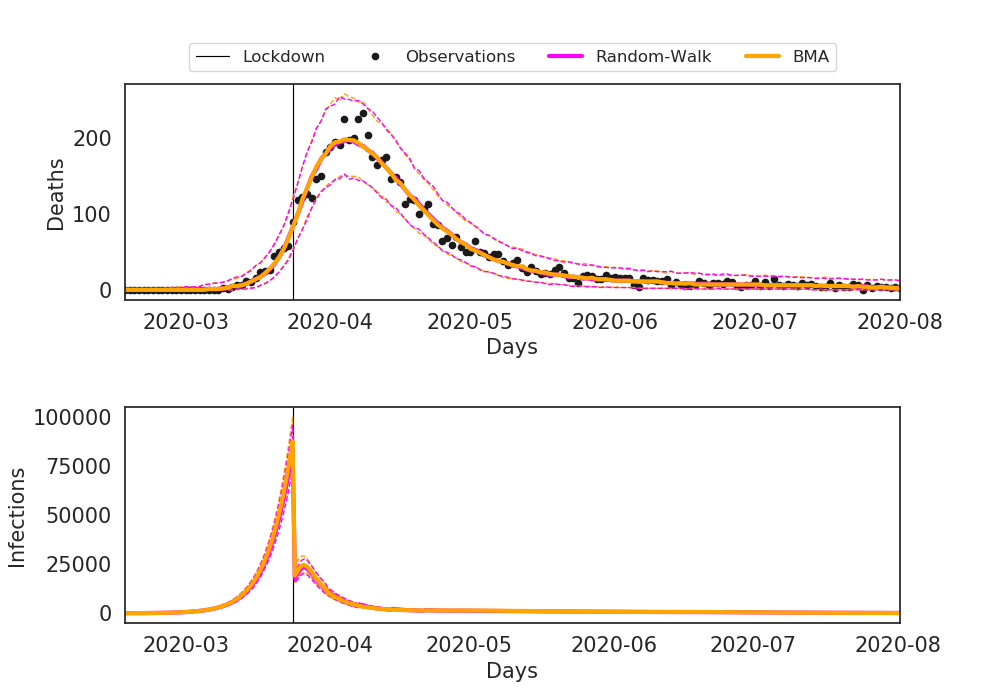}
\caption{Goodness-of-fit of daily death data (a) and the inferred latent infections (b) for the region \textbf{London}. }
\end{figure}

\begin{figure}[!ht]
\includegraphics[width=\textwidth,keepaspectratio=true]{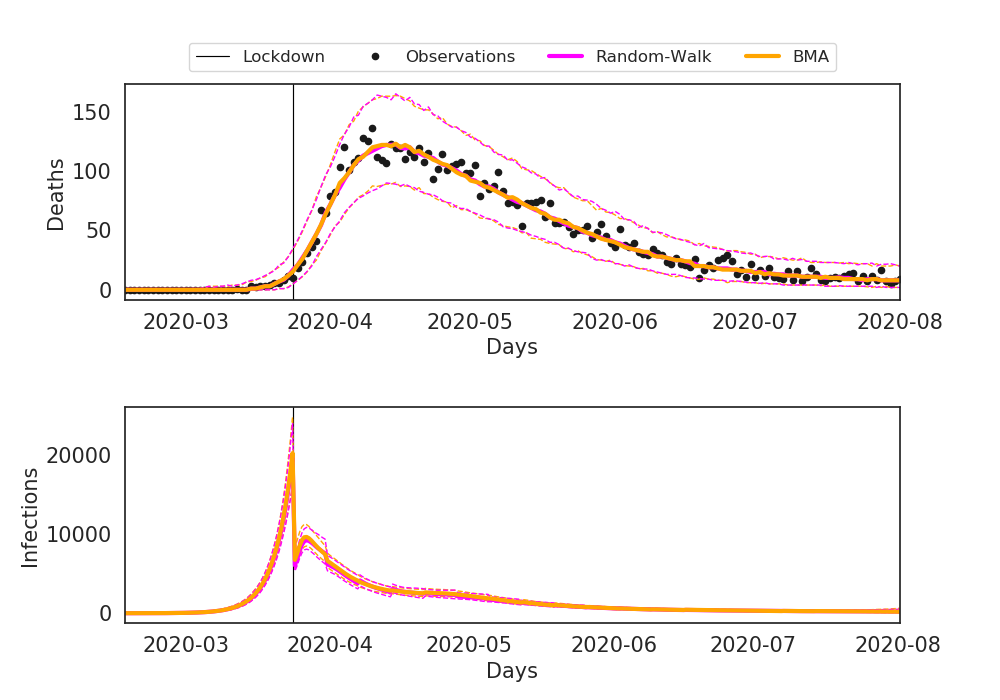}
\caption{Goodness-of-fit of daily death data (a) and the inferred latent infections (b) for the region \textbf{North East and Yorkshire}.}
\end{figure}

\begin{figure}[!ht]
\includegraphics[width=\textwidth,keepaspectratio=true]{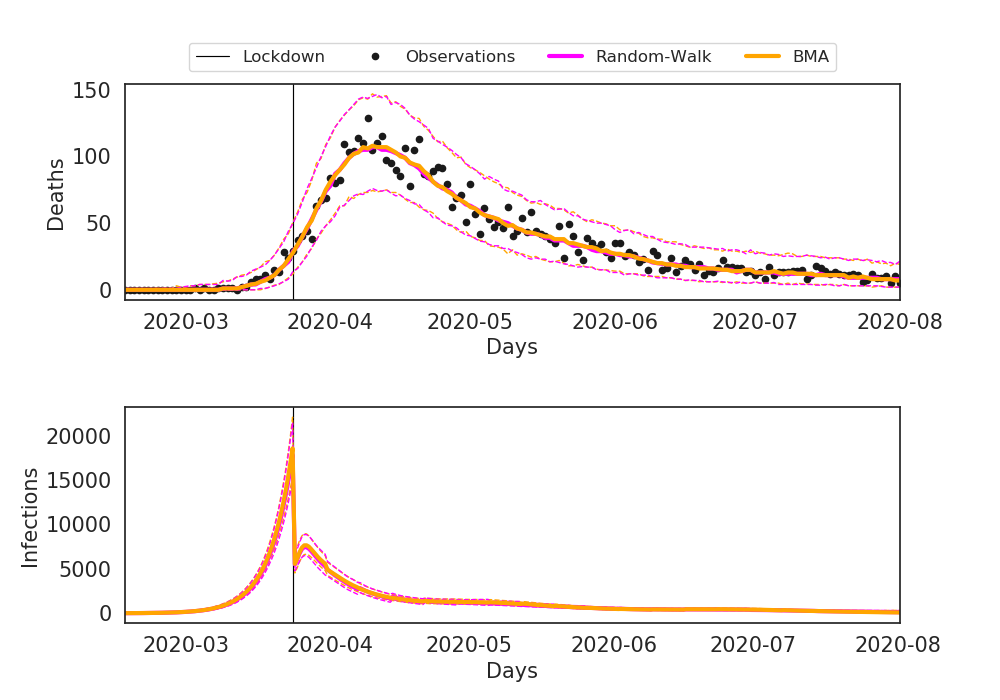}
\caption{Goodness-of-fit of daily death data (a) and the inferred latent infections (b) for the region \textbf{South East}. }
\end{figure}

\end{document}